\documentclass{jfm}

\usepackage{amsmath,amssymb,latexsym, amscd}
\usepackage{exscale, graphics, epsfig, epstopdf, esint, color}
\usepackage{eucal}

\newcommand{\rom}[1]{\uppercase\expandafter{\romannumeral #1\relax}}

\begin{document}

\shorttitle{Stokes waves with constant vorticity}
\shortauthor{S. A. Dyachenko and V. M. Hur}

\title{Stokes waves with constant vorticity:\\ I. numerical computation}

\author{Sergey A. Dyachenko
\corresp{\email{sdyachen@illinois.edu}}
\and Vera Mikyoung Hur}

\affiliation{
Department of Mathematics, University of Illinois at Urbana-Champaign \\ Urbana, IL 61801 USA}

\maketitle

\begin{abstract}
Periodic traveling waves are numerically computed in a constant vorticity flow subject to the force of gravity. The Stokes wave problem is formulated via a conformal mapping as a nonlinear pseudo-differential equation, involving a periodic Hilbert transform for a strip, and solved by the Newton-GMRES method. It works well with a fast Fourier transform and is more effective than a boundary integral method. The result is in excellent agreement, qualitatively and quantitatively, with earlier ones. 

For strong positive vorticity, in the finite or infinite depth, overhanging profiles are found as the steepness increases and tend to a touching wave, whose profile self-intersects somewhere along the trough line, trapping an air bubble; the numerical solutions become unphysical as the steepness increases further and make a gap in the wave speed versus steepness plane; a touching wave then takes over and the physical solutions follow in the wave speed versus steepness plane until they ultimately tend to an extreme wave, which exhibits a sharp corner at the crest. Overhanging waves of nearly maximum heights are found to approach rigid body rotation of a fluid disk as the strength of positive vorticity increases. 
\end{abstract}


\section{Introduction}\label{sec:intro}

\cite{Stokes1847} \citep[see also][]{Stokes1880} made many contributions about periodic waves at the surface of an incompressible inviscid fluid in two dimensions, subject to the force of gravity, traveling a long distance at a practically constant velocity without change of form. For instance, he observed that crests tend to sharpen and troughs flatter as the amplitude increases, and conjectured that the wave of greatest height exhibits a $120^\circ$ corner at the crest. \cite{AFT1982} proved that a limiting wave exists, whose angle at the crest is $120^\circ$. 
In an irrotational flow of infinite depth, Stokes waves are much studied analytically and numerically. 
Some recent advances are based on the formulation of the problem as a nonlinear pseudo-differential equation, involving the periodic Hilbert transform --- namely, the Babenko equation. For instance, \cite{DLK2016, Lushnikov2016, LDS2017} numerically approximated the wave of greatest height and uncovered the structure of the singularities in meticulous detail. 

The zero vorticity assumption may be justified in some situations. Moreover, in the absence of initial vorticity, boundaries or currents, water waves will have zero vorticity at all later times. But rotational effects are significant in many situations. For instance, in any region where wind blows, there is a surface drift of the water, and wave parameters, such as maximum wave height, are sensitive to the velocity at a wind-drift boundary layer. Moreover, currents produce shear at the bed of the sea or a river; see \cite{PTdS1988}, for instance.

For arbitrary vorticity, \cite{CS2004} worked out the global bifurcation of Stokes waves in the finite depth,  \cite{Hur2006, Hur2011} in the infinite depth, and \cite{KS1,KS2} numerically computed, assuming that there is no overhanging or internal stagnation. For zero vorticity, a Stokes wave is necessarily the graph of a single valued function and, moreover, the wave speed exceeds the directional particle velocity inside the fluid. But, even for constant vorticity, \cite{SS1985, PTdS1988, RMN2017}, among others, numerically observed overhanging profiles and interior stagnation points. 

Constant vorticity is of particular interest because of its analytical tractability. 
Moreover, it is representative of a wide range of physical scenarios. When waves are short compared with the vorticity length scale, the vorticity at a surface layer is dominant in the wave dynamics. Moreover, when waves are long compared with the fluid depth, the mean vorticity is more important than its specific distribution; see \cite{PTdS1988}, for instance. Examples include tidal currents --- alternating lateral movements of water associated with the rise and fall of the tide --- where positive or negative constant vorticity suitable for the ebb or flood, respectively; see \cite{CSV2016}, for instance.

Recently, \cite{CSV2016} extended the Babenko equation, to permit constant vorticity and finite depth, and demonstrated the global bifurcation of Stokes waves. Moreover, they conjectured that at the boundary of the solution curve (in a suitable function space), one reaches: either an {\em extreme wave}, which exhibits a sharp corner at the crest and whose profile is single valued or overhanging, or a {\em touching wave}, whose profile self-intersects somewhere along the trough line, trapping an air bubble. 

\cite{SS1985} used a boundary integral method and numerically computed Stokes waves in a constant vorticity flow of infinite depth. They found touching waves, among others, which is higher than the extreme wave for some vorticity. Moreover, they detected a {\em fold} in the wave speed versus steepness plane for some vorticity, which implies non-uniqueness. \cite{PTdS1988} extended the result in the finite depth. \cite{VB1996} located a branch of Stokes waves in the infinite depth, which tend to a closed region of fluid in rigid body rotation at the zero gravity limit. 

Here we use a fast Fourier transform and the Newton-GMRES method, and numerically solve the extension of the Babenko equation, permitting constant vorticity and finite depth. 
The result is in excellent agreement, qualitatively and quantitatively, with those of \cite{SS1985, PTdS1988,VB1996}, among others. 

For negative or weak positive vorticity, in the finite or infinite depth, we learn that single valued profiles tend to an extreme wave as the steepness increases, like the well-known result for zero vorticity. But, for strong positive vorticity, we find that overhanging profiles appear as the steepness increases and tend to a touching wave; the numerical solutions become unphysical as the steepness increases further and make a {\em gap} in the wave speed versus steepness plane. By the way, the numerical method in \cite{SS1985}, for instance, diverges in the gap. A touching wave then takes over and the physical solutions follow along a fold until they ultimately tend to an extreme wave, whose profile seems single valued. Moreover, we find that overhanging waves of nearly maximum heights approach rigid body rotation of a fluid disk as the strength of positive vorticity increases. 

\section{Formulation}\label{sec:formulation}

The water wave problem, in the simplest form, concerns the wave motion at the surface of an incompressible inviscid fluid in two dimensions, lying below a body of air, and acted on by gravity. We assume for simplicity that the density~$=1$. Suppose for definiteness that in Cartesian coordinates, waves propagate in the $x$ direction and gravity acts in the negative $y$ direction. Suppose that the fluid occupies the region, bounded above by a free surface and below by the rigid bed $y=-h$ for some constant $h$ in the range $(0,\infty]$. Let $y=\eta(x;t)$, $x\in\mathbb{R}$, represent the fluid surface at time $t$. We assume for now that $\eta$ is single valued (but see the discussion following \eqref{def:y(u)}). Let 
\[
\Omega(t)=\{(x,y)\in\mathbb{R}^2:-h<y<\eta(x;t)\}\quad\text{and}\quad\Gamma(t)=\{(x,\eta(x;t)):x\in\mathbb{R}\}.
\]

Let $\boldsymbol{u}=\boldsymbol{u}(x,y;t)$ denote the velocity of the fluid at the point $(x,y)$ and time $t$, and $p=p(x,y;t)$ the pressure. They satisfy the Euler equations for an incompressible fluid:
\begin{subequations}\label{E:ww0}
\begin{gather}
\boldsymbol{u}_t+(\boldsymbol{u}\cdot\nabla)\boldsymbol{u}=-\nabla p+(0,-g)\label{E:Euler} 
\intertext{and}
\nabla\cdot\boldsymbol{u}=0\label{E:incomp}
\end{gather}
in $\Omega(t)$, where $g$ is the constant due to gravitational acceleration. Throughout, we express partial differentiation by a subscript, $\nabla=(\partial_x,\partial_y)$ and $\Delta$ the Laplacian. We assume that the vorticity
\begin{equation}
\omega:=\nabla\times\boldsymbol{u}
\end{equation}
is constant. By the way, if vorticity is constant everywhere in the fluid at the initial time then it remains so at all later times, so long as the fluid region is two dimensional and simply connected.   

The kinematic and dynamic conditions at the fluid surface:
\begin{equation}
\eta_t+\boldsymbol{u}\cdot\nabla (\eta-y)=0\quad\text{and}\quad p=p_{atm}\quad\text{at $\Gamma(t)$}
\end{equation}
state, respectively, that the fluid particles do not invade the air, nor vice versa, and that the pressure at the fluid surface equals the constant atmospheric pressure~$=p_{atm}$. Here we assume that the air is quiescent and neglect the effects of surface tension. In the finite depth, where $h<\infty$, the boundary condition at the fluid bed:
\begin{equation}
\boldsymbol{u}\cdot(0,-1)=0\quad\text{at $y=-h$}
\end{equation}
\end{subequations}
states that the fluid particles at the bed remain so at all times. We assume in addition that the solutions of \eqref{E:ww0} are $2L$ periodic in the $x$ variable for some $L$.

For any $\omega\in \mathbb{R}$, $h\in(0,\infty)$ and $c\in\mathbb{R}$, it is straightforward to verify that
\begin{equation}\label{def:shear}
\eta(x;t)=0,\quad\boldsymbol{u}(x,y;t)=(-\omega y-c,0)\quad\text{and}\quad p(x,y;t)=p_{atm}-gy,
\end{equation}
where $x\in\mathbb{R}$ and $y\in(-h,0)$, solve \eqref{E:ww0} at all times. They make a linear shear flow, for which the fluid surface is horizontal, the fluid velocity varies linearly in the $y$ direction, and the pressure is hydrostatic. We assume that some external effects such as wind produce a flow of the kind and restrict the attention to the wave propagation in \eqref{def:shear}. 

Suppose that
\begin{equation}\label{def:Phi}
\boldsymbol{u}(x,y;t)=(-\omega y-c,0)+\nabla\Phi(x,y;t)\quad\text{in $\Omega(t)$},
\end{equation}
whence \eqref{E:incomp} implies that
\[
\Delta\Phi=0\quad\text{in $\Omega(t)$}
\]
at all times. Namely, $\Phi$ is a velocity potential for the irrotational perturbation from \eqref{def:shear}. By the way, for arbitrary vorticity, the perturbation from the shear flow --- not necessarily linear --- becomes rotational, whence $\Phi$ is no longer viable to use. Let $\Psi$ be a harmonic conjugate of $\Phi$. Namely, $\Psi$ is a stream function for the irrotational perturbation from \eqref{def:shear}. Clearly,
\begin{equation}\label{def:Psi}
\boldsymbol{u}=(-\omega y-c,0)+\nabla\times\Psi
\end{equation} 
and $\Delta\Psi=0$ in $\Omega(t)$ at all times. 

We substitute \eqref{def:Phi} and \eqref{def:Psi} into \eqref{E:Euler}, and we make an explicit calculation to arrive at 
\[
\Phi_t+\tfrac12(\Phi_x^2+\Phi_y^2)-(\omega y+c)\Phi_x+\omega\Psi+p-p_{atm}+gy=b(t)
\]
for an arbitrary function $b(t)$. We substitute \eqref{def:Phi} and \eqref{def:Psi} into the other equations of \eqref{E:ww0}, likewise. The result becomes
\begin{subequations}\label{E:ww}
\begin{align} 
&\Delta\Phi=0 &&\text{in $\Omega(t)$},\label{E:ww;Laplace} \\
&\eta_t+(\Phi_x-\omega y-c)\eta_x=\Phi_y &&\text{at $\Gamma(t)$}, \label{E:ww;K}\\
&\Phi_t+\tfrac12|\nabla\Phi|^2-(\omega\eta+c)\Phi_x+\omega\Psi+g\eta=b(t) &&\text{at $\Gamma(t)$}\label{E:ww;B}
\intertext{and}
&\Phi_y=0&&\text{at $y=-h$}.\label{E:ww;bed}
\end{align}
Note that $\eta$ and $\Phi$, $\Psi$ are $2L$ periodic in the $x$ variable.

In the infinite depth, where $h=\infty$, we replace \eqref{E:ww;bed} by
\begin{equation}\label{E:ww;infty}
\Phi,\Psi\to0\quad\text{as $y\to-\infty$}\quad\text{uniformly for $x\in\mathbb{R}$}.
\end{equation}
\end{subequations} 
Moreover, we may assume that $p\to p_{atm}-gy$ as $y\to-\infty$, whence $b(t)=0$. But \eqref{def:Phi} implies that
\[
\boldsymbol{u}\to(-\omega y-c,0)\quad\text{as $y\to-\infty$}.
\]
Therefore, nonzero constant vorticity in the infinite depth seems physically unrealistic. Nevertheless, \eqref{E:ww} makes sense theoretically for any $h\in(0,\infty]$. Moreover, the infinite depth offers an auxiliary conformal mapping for effective numerical computation; see Section \ref{sec:infty-depth} and references therein for details. The effects of depth turn out to change the amplitude of a Stokes wave and other quantities, and they are insignificant otherwise. 
In stark contrast, the effects of constant vorticity are profound on limiting waves and other fundamental issues. 

\subsection{Reformulations via conformal mapping}\label{sec:reformulation}

We reformulate \eqref{E:ww} via a conformal mapping of the fluid region from a strip, or from a half plane in the infinite depth. The idea traces back to \cite{Stokes1880} in the steady wave setting and was explored in the unsteady wave setting by \cite{Ovsyannikov1973} and, later, \cite{MOI1981, Tanveer1991, Tanveer1993, ZDAV2002}, among others. Below, we proceed along the same line as the argument in \cite{DKSZ1996, DZK1996}, but with suitable modifications to accommodate constant vorticity.

In what follows, we identify $\mathbb{R}^2$ with $\mathbb{C}$ whenever it is convenient to do so. 

\subsubsection*{Conformal mapping}

Suppose that
\begin{equation}\label{def:conformal}
z=z(w;t),\quad \text{where}\quad w=u+iv\quad\text{and}\quad z=x+iy,
\end{equation}
conformally maps
\[
\Sigma_d:=\{u+iv\in\mathbb{C}:-d<v<0\}
\]
of $2\upi$ period in the $u$ variable to $\Omega(t)$ of $2L$ period in the $x$ variable at time $t$ for some $d$ in the range $(0,\infty]$. 
Suppose that \eqref{def:conformal} extends to map $\{u+i0:u\in\mathbb{R}\}$ to $\Gamma(t)$ and, moreover, $\{u-id:u\in\mathbb{R}\}$ to $\{x-ih:x\in\mathbb{R}\}$ if $d,h<\infty$, and $-i\infty$ to $-i\infty$ if $d,h=\infty$. Clearly, $x$ and $y$ enjoy the Cauchy-Riemann equations:
\begin{equation}\label{E:CR(x,y)}
x_u=y_v\quad\text{and}\quad x_v=-y_u
\end{equation}
in $\Sigma_d$. Moreover, 
\begin{equation}\label{E:periodicity}
x(u+2\upi+iv;t)=x(u+iv;t)+2L\quad\text{and}\quad y(u+2\upi+iv;t)=v(u+iv;t)
\end{equation}
for $u+iv\in\overline{\Sigma_d}$. 

Therefore, in the finite depth, 
\[
\Delta y=0\quad\text{in $\Sigma_d$}\quad\text{and}\quad y=-h\quad\text{at $v=-d$}.
\]
Suppose that
\begin{equation}\label{E:y(u,0)}
y(u+i0;t)=\sum_{k\in\mathbb{Z}}\widehat{y}(k;t)e^{iku}\quad\text{for $u\in\mathbb{R}$}
\end{equation}
in the Fourier series, where
\[
\widehat{y}(k;t)=\frac{1}{2\upi}\int^{\upi}_{-\upi} y(u+i0;t)e^{iku}~du,
\]
whence
\begin{equation}\label{E:y(u,v)}
y(u+iv;t)=\frac{\widehat{y}(0;t)+h}{d}v+\widehat{y}(0;t)
+\sum_{k\neq0,\in\mathbb{Z}}\frac{\sinh(k(v+d))}{\sinh(kd)}\widehat{y}(k;t)e^{iku}
\end{equation}
for $u+iv\in\overline{\Sigma_d}$. The Cauchy-Riemann equations imply 
\begin{equation}\label{E:x(u,v)}
x(u+iv;t)=\frac{\widehat{y}(0;t)+h}{d}u
-\sum_{k\neq0,\in\mathbb{Z}}i\frac{\cosh(k(v+d))}{\sinh(kd)}\widehat{y}(k;t)e^{iku}
\end{equation}
for $u+iv\in\overline{\Sigma_d}$ up to an additive constant. We infer from \eqref{E:x(u,v)} and \eqref{E:y(u,v)} that 
\[
x_u^2+y_u^2\neq0\quad\text{in $\overline{\Sigma_d}$}.
\] 
Moreover, we infer from \eqref{E:x(u,v)} and the former equation of \eqref{E:periodicity} that
\[
\frac{L}{\upi}=\frac{\widehat{y}(0;t)+h}{d},
\]
which relates the ``mean conformal depth" $d$ and the ``mean fluid depth" $h$ (see the discussion following \eqref{E:mean0(t)}), depending on the solution of \eqref{def:conformal}. 

In what follows, we assume, without loss of generality, that $L=\upi$, whence the above simplifies to
\begin{equation}\label{E:dh}
d=\langle y\rangle+h,
\end{equation}
where 
\begin{equation}\label{def:mean}
\langle f\rangle=\frac{1}{2\upi}\int_{-\upi}^{\upi} f(u)~du
\end{equation}
is the mean over one period of a $2\upi$ periodic function $f$. 
Consequently, \eqref{E:x(u,v)} simplifies to
\begin{equation}\label{E:x(u,0)}
x(u+i0;t)=u-\sum_{k\neq0}i\coth(kd)\widehat{y}(k;t)e^{iku}\quad\text{for $u\in\mathbb{R}$}.
\end{equation}

\subsubsection*{Reformulation via conformal mapping}

Recall \eqref{def:conformal}, and let, by abuse of notation,
\begin{equation}\label{def:y(u)}
(x+iy)(u;t)=(x+iy)(u+i0;t)\quad\text{for $u\in\mathbb{R}$}.
\end{equation}
Therefore,
\[
y(u;t)=\eta(x(u;t);t).
\]
In what follows, we allow that $\eta$ be multi valued. By the way, one may extend \eqref{E:ww} mutatis mutandis when the fluid surface is the trajectory of a parametric curve. A chain rule calculation reveals that
\[
y_u=\eta_xx_u\quad\text{and}\quad y_t=\eta_xx_t+\eta_t.
\]

Recall \eqref{def:Phi} and \eqref{def:Psi}, and let
\begin{equation}\label{def:phi}
(\phi+i\psi)(w;t)=(\Phi+i\Psi)(z(w;t);t)\quad\text{for $w\in\Sigma_d$}.
\end{equation}
Namely, $\phi+i\psi$ is a conformal velocity potential for the irrotational perturbation from \eqref{def:shear}. Since $\Phi+i\Psi$ is holomorphic in $\Omega(t)$ and since $z:\Sigma_d\to\Omega(t)$ is conformal, $\phi$ and $\psi$ enjoy the Cauchy-Riemann equations:
\begin{equation}\label{E:CR(p,q)}
\phi_u=\psi_v\quad\text{and}\quad\phi_v=-\psi_u
\end{equation}
in $\Sigma_d$. A chain rule calculation and \eqref{E:CR(x,y)}, \eqref{E:CR(p,q)} reveal that
\[
\begin{pmatrix}\Phi_x\\ \Phi_y\end{pmatrix}
=\frac{1}{x_uy_u-x_vy_v}\begin{pmatrix}y_v&-y_u\\-x_v&x_u\end{pmatrix}
\begin{pmatrix}\phi_u\\ \phi_v\end{pmatrix}
=\frac{1}{x_u^2+y_u^2}\begin{pmatrix}x_u&-y_u\\ y_u&x_u\end{pmatrix}
\begin{pmatrix}\phi_u\\-\psi_u\end{pmatrix},
\]
where $x_u^2+y_u^2\neq0$ in $\overline{\Sigma_d}$ by \eqref{E:x(u,v)} and \eqref{E:y(u,v)}. Moreover, $\phi_t=\Phi_xx_t+\Phi_yy_t+\Phi_t$. Let, by abuse of notation,
\begin{equation}\label{def:phi(u)}
(\phi+i\psi)(u;t)=(\phi+i\psi)(u+i0;t)\quad\text{for $u\in\mathbb{R}$}.
\end{equation}

We substitute \eqref{def:y(u)} and \eqref{def:phi(u)} into \eqref{E:ww;K} and \eqref{E:ww;B}, and we use the result from the chain rule calculations to arrive at
\begin{subequations}\label{E:ww-conf}
\begin{align}
x_u&y_t-y_ux_t+\psi_u-(\omega y+c)y_u=0\label{E:ww-conf-K}\\
\intertext{and}
\phi_t&-\frac{1}{x_u^2+y_u^2}((x_ux_t+y_uy_t)\phi_u
+(y_ux_t-x_uy_t)\psi_u \label{E:ww-conf-B}\\
&+\tfrac12(\phi_u^2+\psi_u^2)
-(\omega y+c)(x_u\phi_u+y_u\psi_u))+\omega\psi+gy-b(t)=0\notag
\end{align}
at $v=0$. Note that
\begin{equation}\label{E:ww-conf;Laplace}
\Delta y,\Delta\phi=0\quad\text{in $\Sigma_d$}.
\end{equation}
Note that
\begin{equation}\label{E:ww-conf;bed}
y=-h\quad\text{and}\quad\phi_v=0\quad\text{at $v=-d$}\quad\text{if $d$, $h<\infty$}
\end{equation}
by \eqref{E:ww;bed}, and 
\begin{equation}\label{E:ww-conf;infty}
\phi,\psi\to0\quad\text{as $v\to-\infty$}\quad\text{uniformly for $u\in\mathbb{R}$}\quad\text{if $d$, $h=\infty$}
\end{equation} 
\end{subequations}
by \eqref{E:ww;infty} and \eqref{def:phi}. Moreover, $y$ and $\phi$, $\psi$ are $2\upi$ periodic in the $u$ variable. Therefore, \eqref{E:ww-conf} is to rewrite \eqref{E:ww}.

Below, we relate $x$ to $y$ and $\phi$ to $\psi$ at the face of $\Sigma_d$, whereby we reformulate \eqref{E:ww-conf} and, hence, \eqref{E:ww} for $y=y(u;t)$ and $\phi=\phi(u;t)$. It makes use of periodic Hilbert transforms for a strip.

\subsubsection*{Periodic Hilbert transforms for a strip}

For $d$ in the range $(0,\infty)$, let $\mathcal{H}_d$ and $\mathcal{T}_d$ denote Fourier multiplier operators, defined in the periodic setting as
\[
\mathcal{H}_de^{iku}=-i\tanh(kd)e^{iku}\quad \text{for $k\in\mathbb{Z}$}
\]
and
\begin{equation}\label{def:T}
\mathcal{T}_de^{iku}=
\begin{cases}
-i\coth(kd)e^{iku}\quad &\text{if $k\neq0,\in\mathbb{Z}$}, \\
0 &\text{if $k=0$}.
\end{cases}
\end{equation}
Clearly, 
\begin{equation}\label{E:TH=-1}
\mathcal{H}_d\mathcal{T}_d=\mathcal{T}_d\mathcal{H}_d=-1\quad\text{if $k\neq0$}.
\end{equation}
As $d\to\infty$, at least formally, $\mathcal{H}_d$ and $\mathcal{T}_d$ tend to the periodic Hilbert transform, defined likewise as 
\[
\mathcal{H}e^{iku}=-i\,\text{sgn}(k)e^{iku}\quad\text{for $k\in\mathbb{Z}$}.
\]

Among other properties of $\mathcal{H}_d$ and $\mathcal{T}_d$, of particular importance for the present purpose is that the ``Titchmarsh theorem" \citep[see][Theorem~95, for instance]{Titchmarsh} or the Sokhotski-Plemelj theorem \citep[see][for instance]{Plemelj, Gakhov} extends, and $\mathcal{H}_d$ and $\mathcal{T}_d$ relate the real part of a holomorphic and $2\upi$ periodic function in a strip to the imaginary part at the face of the strip, and vice versa. If $F=F(u+iv)$ is holomorphic in the lower half plane of $\mathbb{C}$ and if $F$ vanishes sufficiently rapidly as $v\to-\infty$ then the Titchmarsh theorem states that the real and imaginary parts of $F(\cdot+i0)$ are the Hilbert transforms of each other. For any $d\in(0,\infty)$, likewise, if $F$ is holomorphic in $\Sigma_d$ and $2\upi$ periodic in the $u$ variable and if $\Real F(u+i0)=f(u)$ and $(\Real F)_v(u-id)=0$ for $u\in\mathbb{R}$ then
\begin{equation}\label{E:1-iH}
F(u+i0)=(1-i\mathcal{H}_d)f(u)\quad\text{for $u\in\mathbb{R}$}
\end{equation}
up to an additive imaginary constant. In other words, $1-i\mathcal{H}_d$ makes the face value of a holomorphic and $2\upi$ periodic function in $\Sigma_d$, the normal derivative of whose real part vanishes at the bottom of $\Sigma_d$. Moreover, if $F$ is holomorphic in $\Sigma_d$ and $2\upi$ periodic in the $u$ variable, if $\Imag F(u+i0)=f(u)$ and $\Imag F(u-id)=0$ for $u\in\mathbb{R}$, and if $\langle f\rangle=0$ in addition, where we employ the notation of \eqref{def:mean}, then 
\begin{equation}\label{E:T+i}
F(u+i0)=(\mathcal{T}_d+i)f(u)\quad\text{for $u\in\mathbb{R}$}
\end{equation}
up to an additive real constant. In other words, $\mathcal{T}_d+i$ is the face value of a holomorphic and $2\upi$ periodic function in $\Sigma_d$, whose imaginary part is of mean zero at the face of $\Sigma_d$ and vanishes at the bottom. 

\subsubsection*{Implicit form}

Returning to the water wave problem, in the finite depth, since $\phi+i\psi$ is holomorphic in $\Sigma_d$ and satisfies \eqref{E:ww-conf;bed}, we employ an extension of the Titchmarsh theorem to a strip (see \eqref{E:1-iH}) to show that
\begin{equation}\label{E:Hp}
(\phi+i\psi)(u;t)=(1-i\mathcal{H}_d)\phi(u;t)
\end{equation}
up to an additive imaginary constant. Moreover, we use \eqref{E:y(u,0)}, \eqref{E:x(u,0)} and \eqref{def:T} to show that
\begin{equation}\label{E:Ty}
(x+iy)(u;t)=u+(\mathcal{T}_d+i)y(u;t).
\end{equation}
By the way, an extension of the Titchmarsh theorem (see \eqref{E:T+i}) does not apply to $x+iy$ because $y$ needs not be of mean zero at the face of $\Sigma_d$. (See the discussion following \eqref{E:mean0(t)}.) In the infinite depth, the Titchmarsh theorem implies \eqref{E:Hp} and \eqref{E:Ty}, where the periodic Hilbert transform replaces $\mathcal{H}_d$ and $\mathcal{T}_d$. 

To proceed, in the finite depth, we substitute \eqref{E:Hp} and \eqref{E:Ty} into \eqref{E:ww-conf-K} and \eqref{E:ww-conf-B}, to arrive at
\begin{subequations}\label{E:implicit}
\begin{align}
&(1+\mathcal{T}_dy_u)y_t-y_u\mathcal{T}_dy_t-\mathcal{H}_d\phi_u-(\omega y+c)y_u=0\label{E:implicit-K}
\intertext{and}
&((1+\mathcal{T}_dy_u)^2+y_u^2)(\phi_t+gy-\omega\mathcal{H}_d\phi-b(t)) \notag\\
&\quad-((1+\mathcal{T}_dy_u)\mathcal{T}_dy_t+y_uy_t)\phi_u
+(y_u\mathcal{T}_dy_t-(1+\mathcal{T}_dy_u)y_t)\mathcal{H}_d\phi_u \label{E:implicit-B}\\
&\quad\quad+\tfrac12(\phi_u^2+(\mathcal{H}_d\phi_u)^2)
-(\omega y+c)((1+\mathcal{T}_dy_u)\phi_u-y_u\mathcal{H}_d\phi_u)=0.\notag
\end{align}
\end{subequations}
Note that $y=y(u;t)$ and $\phi=\phi(u;t)$ are $2\upi$ periodic in the $u$ variable. We claim that \eqref{E:implicit} is equivalent to \eqref{E:ww-conf} and, hence, \eqref{E:ww}, provided that $d$ and $h$ are related by \eqref{E:dh}. Indeed, $y$ and $\phi$ extend as the imaginary and real parts of holomorphic and $2\upi$ periodic functions in $\Sigma_d$, which satisfy \eqref{E:ww-conf;bed}. In the infinite depth, \eqref{E:implicit} is equivalent to \eqref{E:ww-conf} and, hence, \eqref{E:ww}, likewise, where the periodic Hilbert transform replaces $\mathcal{H}_d$ and $\mathcal{T}_d$. Moreover, in an irrotational flow, \eqref{E:implicit} agrees with what \cite{DKSZ1996, DZK1996}, for instance, derived. 

We integrate \eqref{E:implicit-K} over the periodic interval $[-\upi,\upi]$ and use that $\mathcal{T}_d$ is anti-self-adjoint, to show that
\begin{equation}\label{E:mean0(t)}
\frac{d}{dt}\langle y(1+\mathcal{T}_dy_u)\rangle=0.
\end{equation}
Therefore, if we locate the coordinates of the fluid region so that $\langle y(1+\mathcal{T}_dy_u)\rangle=0$ at the initial time then it remains so at all later times; $y$ then measures the fluid surface displacement from zero and $h$ the mean fluid depth. In the infinite depth, the periodic Hilbert transform replaces $\mathcal{T}_d$. 

\subsubsection*{Explicit form}

Concluding the reformulations, we solve \eqref{E:implicit} for $y_t$ and $\phi_t$ explicitly.

In the finite depth, since $z$ is holomorphic in $\Sigma_d$ and since $|z_u|^2\neq0$ in $\overline{\Sigma_d}$ by \eqref{E:x(u,v)} and \eqref{E:y(u,v)}, $z_t/z_u$ is holomorphic in $\Sigma_d$. Note that
\begin{equation}\label{E:Im(z_t/z_u)}
\Imag\frac{z_t}{z_u}=\frac{x_uy_t-y_ux_t}{|z_u|^2}
=\frac{\mathcal{H}_d\phi_u+(\omega y+c)y_u}{|z_u|^2}=:\frac{-\chi_u}{|z_u|^2}\quad\text{at $v=0$}
\end{equation}
by \eqref{E:ww-conf-K} and \eqref{E:Hp}, and $\Imag(z_t/z_u)=0$ at $v=-d$ by \eqref{E:ww-conf;bed}.
By the way,
\begin{equation}\label{def:stream}
\chi=\psi-(\tfrac12\omega y^2+cy)
\end{equation}
makes a conformal stream function by  \eqref{def:Psi} and \eqref{def:phi}. Note that $\langle \Imag(z_t/z_u)\rangle=0$ for all $v\in[-d,0]$ by the Cauchy-Riemann equations and \eqref{E:ww-conf;bed}. Therefore, an extension of the Titchmarsh theorem to a strip (see \eqref{E:T+i}) implies that
\[
\frac{z_t}{z_u}=(\mathcal{T}_d+i)\Big(\frac{-\chi_u}{|z_u|^2}\Big)\quad\text{at $v=0$}.
\]
Or, equivalently,
\begin{equation}\label{E:K}
x_t=((1+\mathcal{T}_dy_u)\mathcal{T}_d-y_u)\Big(\frac{-\chi_u}{|z_u|^2}\Big)\quad\text{and}\quad
y_t=(1+\mathcal{T}_dy_u+y_u\mathcal{T}_d)\Big(\frac{-\chi_u}{|z_u|^2}\Big)\quad\text{at $v=0$}.
\end{equation}
Moreover,
\begin{equation}\label{E:Re(z_t/z_u)}
\Real\frac{z_t}{z_u}=\frac{x_ux_t+y_uy_t}{|z_u|^2}=\mathcal{T}_d\Big(\frac{-\chi_u}{|z_u|^2}\Big)
\quad\text{at $v=0$}.
\end{equation} 

We substitute \eqref{E:Im(z_t/z_u)} and \eqref{E:Re(z_t/z_u)} into \eqref{E:ww-conf-B}, to arrive at 
\begin{multline}\label{E:B}
\phi_t+\phi_u\mathcal{T}_d\Big(\frac{\chi_u}{|z_u|^2}\Big)
-\frac{1}{|z_u|^2}(\tfrac12(\phi_u^2-\psi_u^2)(\omega y+c)(1+\mathcal{T}_dy_u)\phi_u)\\
-\omega\mathcal{H}_d\phi+gy-b(t)=0\quad\text{at $v=0$}.
\end{multline}
But an extension of the Titchmarsh theorem to a strip (see \eqref{E:1-iH}) implies that
\[
(\phi_u-i\mathcal{H}_d\phi_u)^2=\phi_u^2-(\mathcal{H}_d\phi_u)^2-2i\phi_u\mathcal{H}_d\phi_u
\]
is the face value of the holomorphic and $2\upi$ periodic function $=(\phi_u+i\psi_u)^2$ in $\Sigma_d$, the normal derivative of whose real part vanishes at the bottom of $\Sigma_d$ by the Cauchy-Riemann equations and  \eqref{E:ww-conf;bed}. It then follows from \eqref{E:1-iH} and \eqref{E:TH=-1} that
\[
\phi_u^2-(\mathcal{H}_d\phi_u)^2=-2\mathcal{T}_d(\phi_u\mathcal{H}_d\phi_u).
\]

We substitute \eqref{def:stream}, \eqref{E:Ty}, \eqref{E:Hp} and the above into the latter equation of \eqref{E:K} and \eqref{E:B}, to arrive at 
\begin{align}\label{E:explicit}
y_t=&(1+\mathcal{T}_dy_u+y_u\mathcal{T}_d)
\Big(\frac{\mathcal{H}_d\phi_u+(\omega y+c)y_u}{(1+\mathcal{T}_dy_u)^2+y_u^2}\Big) \notag
\intertext{and}
\phi_t=&-\phi_u\mathcal{T}_d
\Big(\frac{\mathcal{H}_d\phi_u+(\omega y+c)y_u}{(1+\mathcal{T}_dy_u)^2+y_u^2}\Big) \notag \\
&+\frac{1}{(1+\mathcal{T}_dy_u)^2+y_u^2}(\mathcal{T}_d(\phi_u\mathcal{H}_d\phi_u)
+(\omega y+c)(1+\mathcal{T}_dy_u)\phi_u)+\omega\mathcal{H}_d\phi-gy+b(t).
\end{align}
Note that $y=y(u;t)$ and $\phi=\phi(u;t)$ are $2\upi$ periodic in the $u$ variable. Clearly, \eqref{E:explicit} is equivalent to \eqref{E:ww-conf} and, hence, \eqref{E:implicit}. Therefore, \eqref{E:explicit} is to solve \eqref{E:implicit} for $y_t$ and $\phi_t$ explicitly. In the infinite depth, \eqref{E:explicit} is equivalent to \eqref{E:ww-conf} and, hence, \eqref{E:implicit}, likewise, where the periodic Hilbert transform replaces $\mathcal{H}_d$ and $\mathcal{T}_d$. Moreover, in an irrotational flow, \eqref{E:explicit} agrees with what \cite{DKSZ1996, DZK1996}, for instance, derived. 

\subsection{The Stokes wave problem}\label{sec:Stokes}

We turn the attention to the solutions of \eqref{E:explicit}, for which $y_t$, $\phi_t=0$ and $b(t)=$ constant, and, hence, the steady solutions of \eqref{E:ww}. They make Stokes waves, permitting constant vorticity and finite depth.

In what follows, the prime means ordinary differentiation in the $u$ variable. 

\subsubsection*{Formulation via conformal mapping}

In the finite depth, we substitute $y_t=0$ into the latter equation of \eqref{E:K} to arrive at
\begin{equation}\label{E:psi'}
\psi'=\omega yy'+cy'\quad\text{at $v=0$}.
\end{equation}
Indeed, $(1+\mathcal{T}_dy')^2+(y')^2\neq 0$ pointwise in $\mathbb{R}$ by \eqref{E:Ty} and \eqref{E:x(u,0)}, \eqref{E:y(u,0)}. By the way, \eqref{E:psi'} states that the fluid surface itself makes a streamline. Note from \eqref{E:Hp} and \eqref{E:TH=-1} that
\begin{equation}\label{E:phi'}
\phi'=\mathcal{T}_d(\omega yy'+cy')\quad\text{at $v=0$}.
\end{equation}
Moreover, we substitute $\phi_t=0$ into \eqref{E:B} and use \eqref{E:psi'} and \eqref{E:phi'}, to arrive at
\begin{align*}
(&\mathcal{T}_d(\omega yy'+cy'))^2-(\omega yy'+cy')^2
-2(\omega y+c)(1+\mathcal{T}_dy')\mathcal{T}_d(\omega yy'+cy') \\
&+2\omega((1+\mathcal{T}_dy')^2+(y')^2)(\tfrac12\omega y^2+cy)
-2(b-gy)((1+\mathcal{T}_dy')^2+(y')^2)=0\quad\text{at $v=0$}
\end{align*}
for some constant $b\in\mathbb{R}$. After a lengthy but straightforward calculation, it simplifies to
\begin{equation}\label{E:Stokes}
(c+\omega y(1+\mathcal{T}_dy')-\omega\mathcal{T}_d(yy'))^2=(c^2+2b-2gy)((1+\mathcal{T}_dy')^2+(y')^2).
\end{equation}
Or, equivalently,
\begin{equation}\label{E:Stokes;y}
y=\frac{1}{2g}\Big(c^2+2b-\frac{(c+\omega y(1+\mathcal{T}_dy')-\omega\mathcal{T}_d(yy'))^2}{(1+\mathcal{T}_dy')^2+(y')^2}\Big).
\end{equation}
In the infinite depth, the periodic Hilbert transform replaces $\mathcal{T}_d$. 
Therefore, the Stokes wave problem, permitting constant vorticity and finite depth, is to find $\omega\in\mathbb{R}$, $d\in(0,\infty]$, $b$, $c\in\mathbb{R}$ and a $2\upi$ periodic function $y$, which satisfy \eqref{E:Stokes} or \eqref{E:Stokes;y}. 

In an irrotational flow of infinite depth, we may take $b=0$ (see the discussion following \eqref{E:ww;infty}), whence \eqref{E:Stokes;y} further simplifies to
\[
y=\frac12\frac{c^2}{g}\Big(1-\frac{1}{(1+\mathcal{H}y')^2+(y')^2}\Big).
\]
The result agrees with what \cite{DKSZ1996}, for instance, derived.

\subsubsection*{Reformulation as an equation of Babenko kind}

Unfortunately, \eqref{E:Stokes} or \eqref{E:Stokes;y} is not convenient for numerical computation because one would have to deal with rational functions of $y$. Moreover, \cite{CSV2016} noted that \eqref{E:Stokes} is not suitable for global bifurcation theory because it does not seem to make a compact operator. Below, we proceed along the same line as the argument in \cite{CSV2016} to reformulate \eqref{E:Stokes} as an equation of ``Babenko kind." It makes use of an extension of the Titchmarsh theorem to a strip (see \eqref{E:T+i}) for various quantities.

We begin by arranging \eqref{E:Stokes} as
\begin{align*}
(c-\omega\mathcal{T}_d(yy'))^2+2(c-\omega\mathcal{T}_d(yy'))\omega y(1+\mathcal{T}_dy')
&+\omega^2y^2(1+\mathcal{T}_dy')^2 \\ &=(c^2+2b-2gy)((1+\mathcal{T}_dy')^2+(y')^2),
\end{align*}
and rearranging as
\begin{equation}\label{E:aux1}
\begin{aligned}
(c-\omega\mathcal{T}_d(yy'))^2+2\omega y(c-\omega\mathcal{T}_d(yy'))(&1+\mathcal{T}_dy')
-\omega^2y^2(y')^2 \\ &=(c^2+2b-2gy-\omega^2y^2)((1+\mathcal{T}_dy')^2+(y')^2).
\end{aligned}
\end{equation}
An extension of the Titchmarsh theorem to a strip (see \eqref{E:T+i}) implies that $\mathcal{T}_d(yy')+iyy'$ makes the face value of a holomorphic and $2\upi$ periodic function in $\Sigma_d$, whose imaginary part is of mean zero at the face of $\Sigma_d$ and vanishes at the bottom, and so does
\[
(c-\omega(\mathcal{T}_d(yy')+iyy'))^2
=(c-\omega\mathcal{T}_d(yy'))^2-\omega^2y^2(y')^2-2i(c-\omega\mathcal{T}_d(yy'))\omega yy'.
\]
It then follows from \eqref{E:aux1} that
\begin{equation}\label{E:aux2}
\begin{aligned}
(c^2&+2b-2gy-\omega^2y^2)((1+\mathcal{T}_dy')^2+(y')^2) \\
&-2\omega y(c-\omega\mathcal{T}_d(yy'))(1+\mathcal{T}_dy')-2i\omega y(c-\omega\mathcal{T}_d(yy'))y' \\
=&(c^2+2b-2gy-\omega^2y^2)((1+\mathcal{T}_dy')^2+(y')^2)
-2\omega y(c-\omega\mathcal{T}_d(yy'))(1+\mathcal{T}_dy'+iy') \\
=&((c^2+2b-2gy-\omega^2y^2)(1+\mathcal{T}_dy'-iy')
-2\omega y(c-\omega\mathcal{T}_d(yy')))(1+\mathcal{T}_dy'+iy')
\end{aligned}
\end{equation}
is the face value of a holomorphic and $2\upi$ periodic function in $\Sigma_d$, whose imaginary part is of mean zero at the face of $\Sigma_d$ and vanishes at the bottom.

Note that $1/(1+\mathcal{T}_dy'+iy')$ is the face value of the holomorphic and $2\upi$ periodic function $=1/z_u$ in $\Sigma_d$, whose imaginary part is of mean zero at the face of $\Sigma_d$ and vanishes at the bottom. Indeed, $z$ is holomorphic in $\Sigma_d$, $|z_u|^2\neq0$ in $\overline{\Sigma_d}$ by \eqref{E:x(u,v)} and \eqref{E:y(u,v)} and, moreover, $\langle\Imag(1/z_u)\rangle=0$ for all $v\in[-d,0]$ by the Cauchy-Riemann equations and \eqref{E:ww-conf;bed}. Therefore, it follows from \eqref{E:aux2} that
\[
(c^2+2b-2gy-\omega^2y^2)(1+\mathcal{T}_dy'-iy')-2\omega y(c-\omega\mathcal{T}_d(yy'))
\]
is the face value of a holomorphic and $2\upi$ periodic function in $\Sigma_d$, whose imaginary part is of mean zero at the face of $\Sigma_d$ and vanishes at the bottom. An extension of the Titchmarsh theorem to a strip (see \eqref{E:T+i}) then implies that
\[
(c^2+2b-2gy-\omega^2y^2)(1+\mathcal{T}_dy')-2\omega y(c-\omega\mathcal{T}_d(yy'))
=-\mathcal{T}_d((c^2+2b-2gy-\omega^2y^2)y')
\]
up to an additive real constant. Or, equivalently,
\begin{multline*}
(c^2+2b)\mathcal{T}_dy'-(g+c\omega)y-g(y\mathcal{T}_dy'+\mathcal{T}_d(yy')) \\
-\tfrac12\omega^2(y^2+y^2\mathcal{T}_dy'+\mathcal{T}_d(y^2y')-2y\mathcal{T}_d(yy'))=\mu,
\end{multline*}
say. An integration over the periodic interval $[-\upi,\upi]$ reveals that
\[
\mu=-g\langle y(1+\mathcal{T}_dy')\rangle-c\omega\langle y\rangle-\tfrac12\omega^2\langle y^2\rangle.
\]
Indeed, $\langle \mathcal{T}_df'\rangle=0$ for any function $f$ by \eqref{def:T} and, moreover, since $f\mapsto \mathcal{T}_df'$ is self-adjoint, 
\[
\langle y^2\mathcal{T}_dy'\rangle =\frac{1}{2\upi}\int^{\upi}_{-\upi} y^2\mathcal{T}_dy'~du
=-\frac{1}{2\upi}\int^{\upi}_{-\upi} y\mathcal{T}_d(y^2)'~du=-\langle 2y\mathcal{T}_d(yy')\rangle.
\]
To recapitulate, 
\begin{equation}\label{E:aux3}
\begin{aligned}
(c^2+2b)\mathcal{T}_dy'-(g+c\omega)y-g(y\mathcal{T}_dy'+&\mathcal{T}_d(yy')) \\
-\tfrac12\omega^2(y^2+y^2\mathcal{T}_dy'&+\mathcal{T}_d(y^2y')-2y\mathcal{T}_d(yy')) \\
&+g\langle y(1+\mathcal{T}_dy')\rangle+c\omega\langle y\rangle+\tfrac12\omega^2\langle y^2\rangle=0.
\end{aligned}
\end{equation}
In the infinite depth, the periodic Hilbert transform replaces $\mathcal{T}_d$. 

We emphasize that \eqref{E:aux3} is made up of polynomials of $y$ (involving its derivative and $\mathcal{T}_d$), whence it is straightforward to implement in numerical computation. It is the subject of investigation here. Moreover, \cite{CSV2016} verified that the linearization of \eqref{E:aux3} with respect to $y$ and $b$ is a compact operator in a suitable function space, provided that $c^2+2b-2y>0$ pointwise in $\mathbb{R}$, whereby they established a global bifurcation result.

But for any $\omega\in\mathbb{R}$, $d\in(0,\infty]$ and $b$, $c\in\mathbb{R}$, 
\begin{equation*}\label{E:k>0}
\begin{aligned}
y\mapsto&(c^2+2b)\mathcal{T}_dy'-(g+c\omega)y
-g(y\mathcal{T}_dy'+\mathcal{T}_d(yy'))\\
&-\tfrac12\omega^2(y^2+\mathcal{T}_d(y^2y')+y^2\mathcal{T}_dy'-2y\mathcal{T}_d(yy'))
+g\langle y(1+\mathcal{T}_dy')\rangle+c\omega\langle y\rangle+\tfrac12\omega^2\langle y^2\rangle
\end{aligned}
\end{equation*}
maps $2\upi$ periodic functions to $2\upi$ periodic functions of mean zero, whereas 
\begin{equation*}\label{E:k=0}
y\mapsto (c+\omega y(1+\mathcal{T}_dy')-\omega\mathcal{T}_d(yy'))^2-(c^2+2b-2gy)((1+\mathcal{T}_dy')^2+(y')^2)
\end{equation*}
maps $2\upi$ periodic functions to $2\upi$ periodic functions, not necessarily of mean zero. In other words, \eqref{E:aux3} and \eqref{E:Stokes} agree except a constant. It is because $\mathcal{T}_d+i$ makes the face value of a holomorphic and $2\upi$ periodic function in $\Sigma_d$ merely up to an additive real constant. In order to reconcile loss of information from \eqref{E:Stokes} to \eqref{E:aux3}, we require that the solutions of \eqref{E:aux3} in addition satisfy
\begin{equation}\label{E:aux4}
\langle(c+\omega y(1+\mathcal{T}_dy')-\omega\mathcal{T}_d(yy'))^2\rangle
=\langle (c^2+2b-2gy)((1+\mathcal{T}_dy')^2+(y')^2)\rangle.
\end{equation}
Consequently, \eqref{E:aux3} and \eqref{E:aux4} are equivalent to \eqref{E:Stokes}. 

Furthermore, \eqref{E:Stokes} cannot uniquely determine $y$ and $b$, and neither can \eqref{E:aux3} and \eqref{E:aux4}. It is because \eqref{E:ww} is a free boundary problem. Indeed, $b$ depends on the location of the coordinates of the fluid region, among others. Recall \eqref{E:mean0(t)}, and we assume, without loss of generality, that
\begin{equation}\label{E:mean0}
\langle y(1+\mathcal{T}_dy')\rangle=0;
\end{equation}
$y$ then measures the fluid surface displacement from zero and $h$ the mean fluid depth. It in turn simplifies \eqref{E:aux3}.

We assume in addition that the solutions of \eqref{E:aux3}, \eqref{E:aux4} and \eqref{E:mean0} are even. Indeed, 
for arbitrary vorticity, under some assumptions, \cite{Hur2007} and \cite{CEW2007}, among others, proved that a Stokes wave is a priori symmetric about the crest. 

To summarize, the Stokes wave problem, permitting constant vorticity and finite depth, is to find a vorticity $\omega\in\mathbb{R}$, a mean conformal depth $d\in(0,\infty]$, a ``Bernoulli constant" $b\in\mathbb{R}$, a wave speed $c\in\mathbb{R}$, and a $2\upi$ periodic and even function $y$, measuring the fluid surface displacement from zero, which satisfy
\begin{subequations}\label{E:main}
\begin{equation}\label{E:Babenko}
\begin{aligned}
(c^2+2b)\mathcal{T}_dy'&-(g+c\omega)y-g(y\mathcal{T}_dy'+\mathcal{T}_d(yy'))\\
&-\tfrac12\omega^2(y^2+\mathcal{T}_d(y^2y')+y^2\mathcal{T}_dy'-2y\mathcal{T}_d(yy'))
+c\omega\langle y\rangle+\tfrac12\omega^2\langle y^2\rangle=0
\end{aligned}
\end{equation}
and
\begin{equation}\label{E:<Stokes>}
\langle(c+\omega y(1+\mathcal{T}_dy')-\omega(\mathcal{T}_d(yy'))^2\rangle
-\langle (c^2+2b-2gy)((1+\mathcal{T}_dy')^2+(y')^2)\rangle=0.
\end{equation}
\end{subequations}
We usually regard $\omega$ and $d$ as prescribed, and $y$ and $b$ as the unknowns, depending on the parameter $c$, although we at times switch the roles of $c$ and $\omega$ or $d$; see Section~\ref{sec:initial guess}, Section~\ref{sec:max omega} and Section~\ref{sec:min depth}, for instance. In the finite depth, we determine the mean fluid depth $h$ by solving \eqref{E:dh}, depending on the solution. One may instead fix $h$ and determine $d$ as part of the solution. 

We remark that \cite{CSV2016} focused on steady waves to discover \eqref{E:main}, and here we begin by deriving the governing equations in the unsteady wave setting and rediscover \eqref{E:main} by seeking the steady solutions, which is potentially useful for addressing stability and other unsteady wave phenomena. It is a subject of future investigation. Moreover,  \cite{CSV2016} required $\langle y\rangle=0$ in place of \eqref{E:mean0}. But we infer from \eqref{E:mean0(t)} that \eqref{E:mean0} is more suitable for studying unsteady waves. 

In an irrotational flow, \eqref{E:Babenko} simplifies to
\begin{equation}\label{E:Babenko0d}
(c^2+2b)\mathcal{T}_dy'-gy-g(y\mathcal{T}_dy'+\mathcal{T}_d(yy'))=0.
\end{equation}
Moreover, we may redefine the square of the wave speed~$=c^2+2b$ so long as it is positive. Therefore, for zero vorticity, the Stokes wave problem is to find $d\in(0,\infty]$, $c^2+2b\in(0,\infty)$ and a $2\upi$ periodic and even function $y$, which satisfy \eqref{E:Babenko0d}. The result agrees with what \cite{DKSZ1996, DZK1996}, for instance, derived. In stark contrast, for nonzero constant vorticity, one must determine $b$ as part of the solution by solving \eqref{E:Babenko} and \eqref{E:<Stokes>} simultaneously for $y$ and $b$. 

In the infinite depth, in addition, the periodic Hilbert transform replaces $\mathcal{T}_d$ and we may take $b=0$ (see the discussion following \eqref{E:ww;infty}), whence \eqref{E:Babenko0d} further simplifies to
\begin{equation}\label{E:Babenko0}
c^2\mathcal{H}y'-gy-g(y\mathcal{H}y'+\mathcal{H}(yy'))=0.
\end{equation}
\cite{LH1978} proposed a collection of infinitely many equations for the Fourier coefficients of a Stokes wave, which \cite{Babenko1987} rediscovered in the form of \eqref{E:Babenko0}, and, independently, \cite{Plotnikov1992, DKSZ1996, BDT2000a, BDT2000b}, among others. One may regard \eqref{E:main} as to extending the Babenko equation to permit constant vorticity and finite depth. 

We compare \eqref{E:Babenko} and \eqref{E:Babenko0d} to learn that nonzero constant vorticity adds higher order nonlinearities to the equation, whence it may contribute to new wave phenomena. In stark contrast, we compare \eqref{E:Babenko0d} and \eqref{E:Babenko0} to learn that the mean conformal depth merely changes the scaling factor of the Fourier multiplier in the equation, whence it would not influence the qualitative properties of the solutions. The result from the present numerical computation bears it out.

\section{Numerical method}\label{sec:numerical}

We begin by writing \eqref{E:main} in the operator form as 
\begin{equation}\label{E:F=0}
F(y,b;c,\omega,d)=0,
\end{equation}
where $F(y,b;c,\omega,d)=(Y,B)(y,b;c,\omega,d)$,
\begin{subequations}
\begin{align}
Y(y,b;c,\omega,d)=&(c^2+2b)\mathcal{T}_dy'-(g+c\omega)y-g(y\mathcal{T}_dy'+\mathcal{T}_d(yy'))\label{def:Y}\\
&-\tfrac12\omega^2(y^2+\mathcal{T}_d(y^2y')+y^2\mathcal{T}_dy'-2y\mathcal{T}_d(yy'))
+c\omega\langle y\rangle+\tfrac12\omega^2\langle y^2\rangle \notag
\intertext{and}
B(y,b;c,\omega,d)=&\langle(c+\omega y(1+\mathcal{T}_dy')-\omega\mathcal{T}_d(yy'))^2\rangle 
-\langle(c^2+2b-2gy)((1+\mathcal{T}_dy')^2+(y')^2)\rangle.\label{def:B}
\end{align}
\end{subequations}
For any $b$, $c$, $\omega\in\mathbb{R}$ and $d\in(0,\infty]$, $Y(\cdot,b;c,\omega,d)$ maps $2\upi$ periodic and even functions to $2\upi$ periodic and even functions, whence
\[
Y(y,b;c,\omega,d)(u)=\sum_{k\in\mathbb{Z}}\widehat{Y}(k)(y,b;c,\omega,d)e^{iku}\quad\text{for $u\in\mathbb{R}$}
\]
in the Fourier series, where
\begin{subequations}\label{def:Y}
\begin{equation}\label{def:Y1}
\begin{aligned}
\widehat{Y}(k)(y,b;c,\omega,d)=\frac{1}{2\upi}\int^{\upi}_{-\upi} ((c^2&+2b)\mathcal{T}_dy'-(g+c\omega)y-g(y\mathcal{T}_dy'+\mathcal{T}_d(yy')) \\
&-\tfrac12\omega^2(y^2+\mathcal{T}_d(y^2y')+y^2\mathcal{T}_dy'-2y\mathcal{T}_d(yy')))e^{iku}~du
\end{aligned}
\end{equation}
and $\widehat{Y}(k)=\widehat{Y}(-k)$ for all $k\in\mathbb{Z}$. In what follows, we identify $Y$ with $(\widehat{Y}(0),\widehat{Y}(1),\widehat{Y}(2),\dots)$. Note that 
\begin{equation}\label{def:Y0}
\widehat{Y}(0)(y,b;c,\omega,d)=\langle y(1+\mathcal{T}_dy')\rangle.
\end{equation}
\end{subequations}

\subsection{Newton-GMRES method}\label{sec:Newton}

Suppose that 
\begin{equation}\label{def:y(n)}
y^{(n+1)}=y^{(n)}+\delta y^{(n)}\quad\text{and}\quad b^{(n+1)}=b^{(n)}+\delta b^{(n)}
\quad\text{for $n=0,1,2,\dots$}
\end{equation}
solve \eqref{E:F=0} and \eqref{def:Y}-\eqref{def:B} iteratively by the Newton method, where $y^{(0)}$ and $b^{(0)}$ make an initial guess, to be supplied (see Section \ref{sec:initial guess} for details), $\delta y^{(n)}$ and $\delta b^{(n)}$ solve 
\begin{equation}\label{E:dF=-F}
\delta F(y^{(n)},b^{(n)};c,\omega,d)(\delta y^{(n)},\delta b^{(n)})=-F(y^{(n)},b^{(n)};c,\omega,d),
\end{equation}
$F(y^{(n)},b^{(n)};c,\omega,d)$ is defined in \eqref{def:Y}-\eqref{def:B}, and $\delta F(y^{(n)}, b^{(n)};c,\omega,d)$ is the linearization of $F(y,b;c,\omega,d)$ with respect to $y$ and $b$, and evaluated at $y=y^{(n)}$ and $b=b^{(n)}$. We use \eqref{def:Y} and \eqref{def:B}, and we make an explicit calculation to show that 
\[
\delta F(y,b;c,\omega,d)(\delta y,\delta b)
=(\delta\widehat{Y}(0),\delta\widehat{Y}(1),\delta\widehat{Y}(2),\dots,\delta B)(y,b;c,\omega,d)(\delta y,\delta b),
\]
where 
\begin{subequations}\label{def:dF}
\begin{align}
\delta\widehat{Y}(k)(y,b)(\delta y,\delta b)
=&\frac{1}{2\upi}\int^{\upi}_{-\upi} 
((c^2+2b)\mathcal{T}_d(\delta y)'+2\delta b\mathcal{T}_dy'-(g+c\omega)\delta y \notag \\
&\qquad\quad-g(\delta y\mathcal{T}_dy'+y\mathcal{T}_d(\delta y)'+\mathcal{T}_d(y\delta y)') \notag \\
&\qquad\quad-\tfrac12\omega^2(2y\delta y+\mathcal{T}_d(y^2\delta y)'
-[2y\delta y,y]+[y^2,\delta y]))e^{iku}~du\label{def:dY1} 
\intertext{for $k=1,2,\dots$, $[f_1,f_2]:=f_1\mathcal{T}_df_2'-f_2\mathcal{T}_df_1'$,}
\delta\widehat{Y}(0)(y,b)(\delta y,\delta b)
=&\langle\delta y+2y\mathcal{T}_d(\delta y)'\rangle\label{def:dY0} 
\intertext{and}
\delta B(y,b)(\delta y,\delta b)
=&2\omega\langle(c+\omega y(1+\mathcal{T}_dy')-\omega\mathcal{T}_d(yy')) \notag \\
&\qquad\times(\delta y(1+\mathcal{T}_dy')+y\mathcal{T}_d(\delta y)'-\mathcal{T}_d(y\delta y)')\rangle \notag\\
&-2\langle(\delta b-g\delta y)((1+\mathcal{T}_dy')^2+(y')^2)\rangle \notag \\
&-2\langle (c^2+2b-2gy)((1+\mathcal{T}_dy')\mathcal{T}_d(\delta y)'+y'(\delta y)')\rangle.\label{def:dB}
\end{align}
\end{subequations}

We approximate $y^{(n)}$ by a truncated Fourier series and, by abuse of notation, let 
\begin{equation}\label{def:f(u);N}
y^{(n)}(u):=\sum_{k=-N/2}^{N/2-1}\widehat{y^{(n)}}(k)e^{iku}
\end{equation}
for some even $N$, where $\widehat{y^{(n)}}(k)=\widehat{y^{(n)}}(-k)$ for all $k\in\mathbb{Z}$, by symmetry. We approximate $\widehat{y^{(n)}}(k)$ by a discrete Fourier transform and, by abuse of notation, let
\begin{equation}\label{def:f(k);N}
\widehat{y^{(n)}}(k):=\frac{1}{N}\sum_{j=0}^{N-1}y^{(n)}(u_j)e^{iku_j}\quad\text{for $k=-N/2,\dots, N/2-1$},
\end{equation}
where
\begin{equation}\label{def:uj}
u_j=-\upi+2\upi j/N\quad\text{for $j=0,1,\dots,N-1$}
\end{equation}
make uniform grid points of the periodic interval $[-\upi,\upi]$, and $y^{(n)}(u_j)=y^{(n)}(u_{N-j})$ for $j=0,1,\dots,N/2-1$, by symmetry. We compute \eqref{def:f(k);N} using a fast Fourier transform (FFT). We numerically approximate $\mathcal{T}_d(y^{(n)})'$ (see \eqref{def:T}) and polynomial nonlinearities, e.g. $y^{(n)}\mathcal{T}_d(y^{(n)})'$, likewise, using \eqref{def:f(u);N}-\eqref{def:uj}, an FFT and the inverse FFT. 

Together, we numerically approximate \eqref{E:dF=-F}, using \eqref{def:f(u);N}-\eqref{def:uj}, an FFT and the inverse, by
\begin{equation}\label{E:Ax=b}
\begin{aligned}
(\delta\widehat{Y}(0),\delta\widehat{Y}(1),\dots,\delta\widehat{Y}(N/2-1),\delta B)
(y^{(n)},b^{(n)};c,\omega,d)&(\delta y^{(n)},\delta b^{(n)}) \\
=-(\widehat{Y}(0),\widehat{Y}(1),\dots,\widehat{Y}(N/2-1),&B)(y^{(n)},b^{(n)};c,\omega,d),
\end{aligned}
\end{equation}
where $\widehat{Y}(0)$, $\widehat{Y}(1)$, \dots, $\widehat{Y}(N/2-1)$, $B$ are in \eqref{def:Y}-\eqref{def:B}, and $\delta\widehat{Y}(0)$, $\delta\widehat{Y}(1)$, \dots, $\delta\widehat{Y}(N/2-1)$, $\delta B$ in \eqref{def:dF}; $y^{(n)}$ is in \eqref{def:f(u);N} and $\delta y^{(n)}$, likewise; $N$ is the number of the Fourier coefficients or, alternatively, the number of the grid points in \eqref{def:f(u);N}-\eqref{def:uj}. By the way, $(\delta\widehat{Y}(0),\delta\widehat{Y}(1),\dots,\delta\widehat{Y}(N/2-1),\delta B)(y^{(n)},b^{(n)};c,\omega,d)$ is not given explicitly, but for any $(\delta y^{(n)},\delta b^{(n)})$, the left side of \eqref{E:Ax=b} may be computed using an FFT and a pseudo-spectral method. 

It is reasonable to solve \eqref{E:Ax=b} by a Krylov subspace method. Some excellent surveys include \cite{Greenbaum1997, Meurant1999, Saad2003, SS2007}. 
But the conjugate gradient (CG) method does not seem to converge for strong positive vorticity, among others. The conjugate residual (CR) and minimal residual (MINRES) methods are better but unreliable. Perhaps, it is because for any $c$, $\omega \in\mathbb{R}$ and $d\in(0,\infty]$, 
\begin{align*}
(\delta y,\delta b)
\mapsto &(c^2+2b)\mathcal{T}_d(\delta y)'+2\delta b\mathcal{T}_dy'-(g+c\omega)\delta y \\
&-g(\delta y\mathcal{T}_dy'+y\mathcal{T}_d(\delta y)'+\mathcal{T}_d(y\delta y)')
-\tfrac12\omega^2(2y\delta y+\mathcal{T}_d(y^2\delta y)'-[2y\delta y,y]+[y^2,\delta y]),
\end{align*}
where $[f_1,f_2]=f_1\mathcal{T}_df_2'-f_2\mathcal{T}_df_1'$, and, hence, \eqref{def:dY1} are not self-adjoint. 
In stark contrast, for any $b\in\mathbb{R}$, for any $c$, $\omega\in\mathbb{R}$ and $d\in(0,\infty]$, 
\begin{align*}
\delta y\mapsto &(c^2+2b)\mathcal{T}_d(\delta y)'-(g+c\omega)\delta y \\
&-g(\delta y\mathcal{T}_dy'+y\mathcal{T}_d(\delta y)'+\mathcal{T}_d(y\delta y)')
-\tfrac12\omega^2(2y\delta y+\mathcal{T}_d(y^2\delta y)'-[2y\delta y,y]+[y^2,\delta y])
\end{align*}
is self-adjoint. Indeed, in an irrotational flow of infinite depth, where we may take $b=0$, the CG or CR method does converge; see \cite{DLK2016, Lushnikov2016, LDS2017}, for instance. 

For nonzero constant vorticity, the generalized minimal residual (GMRES) method \citep[see][for instance]{SS1986} turns out to converge in the least number of iterates, compared with the CG, CR and MINRES methods. The result reported herein is based on the GMRES method. By the way, one may instead attempt to solve
\[
(\delta F)^*(\delta F)(y^{(n)},b^{(n)};c,\omega,d)(\delta y^{(n)},\delta b^{(n)})
=-(\delta F)^*F(y^{(n)},b^{(n)};c,\omega,d)
\]
by the CG or CR method, where the asterisk denotes the adjoint. It is presently under investigation.

We emphasize that an FFT computes a discrete Fourier transform in $O(N\log N)$ operations, where $N$ is the number of the Fourier coefficients or, alternatively, the number of the grid points (see \eqref{def:f(u);N}-\eqref{def:uj}). To compare, a boundary integral method in \cite{SS1985, PTdS1988}, for instance, would compute a numerical solution in $O(N^2)$ operations, although a customized $\sqrt{N}$  grid points would possibly improve it to $O(N)$. 

A uniform grid is clearly not very effective for nearly extreme waves, whose crests tend to sharpen and troughs flatter. In an irrotational flow of infinite depth, \cite{LDS2017}, for instance, proposed an auxiliary conformal mapping, which adapts the numerical points for high curvature so that an FFT computes a discrete Fourier transform in $O(\sqrt{N}\log N)$ operations in the auxiliary conformal variable for comparable resolution. In Section~\ref{sec:infty-depth}, we exercise the idea for nonzero constant vorticity in the infinite depth. 

\subsection{Initial guess}\label{sec:initial guess}

It turns out that we must supply $y^{(0)}$ and $b^{(0)}$ sufficiently close to a true solution of \eqref{E:F=0} and  \eqref{def:Y}-\eqref{def:B}, in order for the Newton-GMRES method in the previous subsection to converge.

For $\omega\in\mathbb{R}$ and $d\in(0,\infty]$ prescribed, we begin by taking $\omega=0$, $d=\infty$ and a Pad\'e approximation of a Stokes wave, in practice, of small amplitude:
\[
z(w)\sim w+iy_0+\sum_{m=1}^M\frac{\gamma_m}{\tan(w/2)-i\beta_m}\quad\text{for $w=u+iv\in\mathbb{C}$},
\]
where $z(u+iv)$ for $v<0$ is the conformal mapping from the lower half plane of $\mathbb{C}$ to the fluid region, $z(u+i0)$ is the fluid surface (see \eqref{def:conformal}), and $z(u+iv)$ for $v>0$ is an analytic continuation of the conformal mapping to the upper half plane of $\mathbb{C}$; $\beta_m$ and $\gamma_m$ for $m=1,2,\dots,M$ for some $M$ are the poles and residues of the Pad\'e approximation, and $y_0$ is chosen so that \eqref{E:mean0} holds. 
See \cite{DLK2016}, for instance, for details. By the way, $z$ is holomorphic in the lower half plane of $\mathbb{C}$ but it may admit singularities in the upper half plane. A library of $\beta_m$ and $\gamma_m$ from zero to a nearly maximum amplitude is found in \cite{DLK}, for instance, which enables one to reconstruct Stokes waves, in an irrotational flow of infinite depth, for the relative error less than $10^{-26}$. 

We then continue the numerical solution along in $\omega$ and $d$, taking the prior convergent solution as the initial guess and solving \eqref{E:F=0} and \eqref{def:Y}-\eqref{def:B} by the method in the previous subsection, until we reach a solution for the desired values of $\omega$ and $d$. One may instead recall the Stokes wave expansion; see \cite{SS1985, PTdS1988}, for instance. 

To proceed, we fix $\omega$ and $d$, and continue the numerical solution along in $c$, taking the prior convergent solution as the initial guess and solving \eqref{E:F=0} and \eqref{def:Y}-\eqref{def:B} by the method in Section~\ref{sec:Newton}. But there is a caveat. For strong positive vorticity, the solution curve in the wave speed versus amplitude plane turns out to experience a ``gap" of unphysical solutions; see Figure~\ref{fig:c(s);+vor;d=1} and Figure~\ref{fig:c(s);vor=2.25}, for instance. For stronger positive vorticity, the maximum wave speed in the gap becomes considerably larger and, hence, the number of steps in $c$ one would have to take to continue along and trace out the gap. For instance, for $\omega=2.25$ and $d=1$ (see Figure~\ref{fig:c(s);vor=2.25}), the wave speed reaches the order of hundreds in the gap. It is then more effective to continue the physical solution along in $\omega$ (and $c$) from a smaller value of $\omega$.

\subsection{Convergence}\label{sec:convergence}

For a numerical solution $y^{(n)}$ and $b^{(n)}$ of \eqref{E:F=0} and \eqref{def:Y}-\eqref{def:B}, we define the residual as
\[
\text{res}(y^{(n)},b^{(n)})=\sum_{k=-N/2}^{N/2-1}|\widehat{Y}(k)(y^{(n)},b^{(n)})|^2+|B(y^{(n)},b^{(n)})|^2,
\]
where $N$ is the number of the Fourier coefficients in the numerical approximation (see \eqref{def:f(u);N}-\eqref{def:uj}). It measures how far $y^{(n)}$ and $b^{(n)}$ are from a true solution of \eqref{E:F=0} and \eqref{def:Y}-\eqref{def:B}. We say that the Newton method converges if 
\begin{equation}\label{E:res13}
\text{res}(y^{(n)},b^{(n)})<\sqrt{N}10^{-13}.
\end{equation}
But if the wave speed of a numerical solution is of the order of hundreds (see Figure~\ref{fig:c(s);vor=2.25}, for instance), we necessarily relax \eqref{E:res13} to 
\begin{equation}\label{E:res9}
\text{res}(y^{(n)},b^{(n)})<\sqrt{N}10^{-9}.
\end{equation}

For a numerical solution $\delta y^{(n)}$ and $\delta b^{(n)}$ of \eqref{E:dF=-F}-\eqref{def:dF} or, equivalently, \eqref{E:Ax=b}, we define the residual likewise as
\begin{align*}
\delta\text{res}(\delta y^{(n)},\delta b^{(n)})
=&\sum_{k=-N/2}^{N/2-1}|\delta\widehat{Y}(k)(y^{(n)},b^{(n)})(\delta y^{(n)},\delta b^{(n)})
+\widehat{Y}(k)(y^{(n)},b^{(n)})|^2 \\
&+|\delta B(y^{(n)},b^{(n)})(\delta y^{(n)},\delta b^{(n)})+B(y^{(n)},b^{(n)})|^2.
\end{align*}
Because \eqref{E:dF=-F}-\eqref{def:dF} is to better approximate the numerical solutions of \eqref{E:F=0} and \eqref{def:Y}-\eqref{def:B}, we do not have to insist the residual of \eqref{E:dF=-F}-\eqref{def:dF} smaller than a fraction of that of \eqref{E:F=0} and \eqref{def:Y}-\eqref{def:B}. 
As \cite{Yang2010}, for instance, suggested, we say that the GMRES method converges if
\begin{equation}\label{E:res-delta}
\delta\text{res}(\delta y^{(n)},\delta b^{(n)})<\epsilon\,\text{res}(y^{(n)},b^{(n)})
\end{equation}
for some small $\epsilon$, say, $0.01$. 
To compare, \cite{SS1985} and \cite{VB1996} required the residuals less than $10^{-11}$, and \cite{PTdS1988} the residuals less than $10^{-10}$.  

In the present computation, the number of the Newton iterates remains constant as $N$ increases from $256$ to $2^{16}=65536$ and it rarely takes more than $12$, in order for a numerical solution of \eqref{E:F=0} and \eqref{def:Y}-\eqref{def:B} to satisfy \eqref{E:res13}. The number of the GMRES iterates, on the other hand, varies from the order of tens to thousands for $N$ in the range, in order for a numerical solution of \eqref{E:dF=-F}-\eqref{def:dF} to satisfy \eqref{E:res-delta}. 

\subsection{Error}\label{sec:error}

Lastly, we require that a numerical solution $y^{(n)}$ (and $b^{(n)}$) of \eqref{E:F=0} and \eqref{def:Y}-\eqref{def:B} satisfies
\begin{equation}\label{E:error}
|\widehat{y^{(n)}}(N/2)|<10^{-12},
\end{equation}
where $N$ is the number of the Fourier coefficients in the numerical approximation, so that the truncation error in \eqref{def:f(u);N} is insignificant. In general, we have to take $N$ considerably larger for higher waves, in order to accurately resolve the numerical solutions. Indeed, in an irrotational flow of infinite depth, if one requires that a numerical solution satisfies 
\[
|\widehat{y^{(n)}}(N/2)|<N^{-1/2}10^{-26}
\] 
in place of \eqref{E:error}, then $N=256$ for the steepness $=0.0994457$, but $N=2^{27}\approx1.3\times10^8$ in order to estimate the steepness of the wave of greatest height up to $32$ digits; see \cite{DLK2016}, for instance. In the present computation, we take $N$ in the range of $256$ to $2^{16}=65536$, and we do not attempt to find a solution if it requires more than $2^{16}$ Fourier coefficients to satisfy \eqref{E:error}, because it takes too much time. 

\section{Result}\label{sec:result}

If $y$, $b$ and $c$ make a solution of \eqref{E:main} or, equivalently, \eqref{E:F=0} and \eqref{def:Y}-\eqref{def:B} for some $\omega$ and $d$, then it is straightforward to verify that so do $y$, $b$ and $-c$ for $-\omega$ and $d$. In what follows, we assume that $c$ is positive and, in turn, allow that $\omega$ be either positive or negative, representative of waves propagating upstream or downstream, respectively; see \cite{PTdS1988}, for instance. 

We take $g=1$ for simplicity. Recall that the period is $2\pi$. The steepness $s$ measures the crest-to-trough height divided by $2\pi$.

\subsection{Finite depth}\label{sec:finite depth}

Throughout the subsection, $d=1$ for simplicity. 

\subsubsection*{Zero or negative vorticity}

\begin{figure}
\centerline{\includegraphics[scale=1.1]{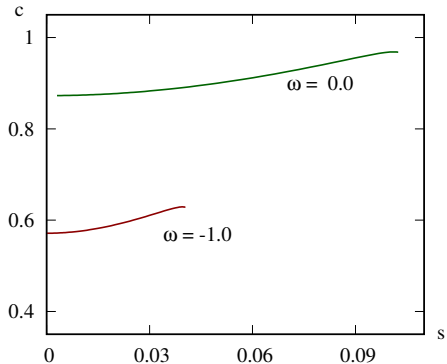}}
\caption{Wave speed versus steepness for $d=1$, $\omega=0$ (green) and $\omega=-1$ (red).}
\label{fig:c(s);vor=0,-1}
\end{figure}

\begin{figure}
\centerline{\includegraphics[scale=1.1]{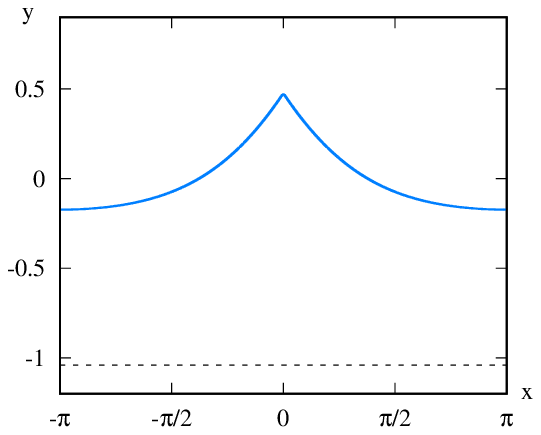}\hspace*{-20pt}
\includegraphics[scale=1.05]{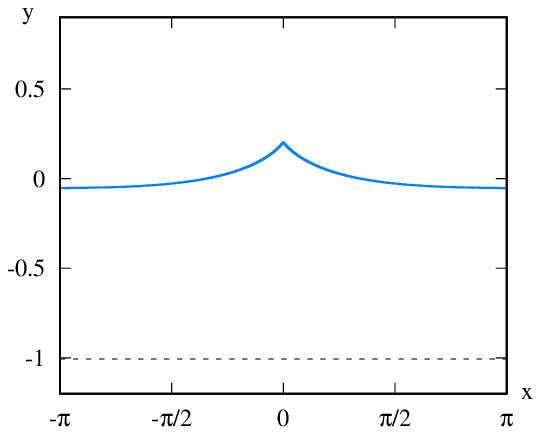}}
\caption{(left) For $\omega=0$ and $d=1$, the wave profile in the $(x,y)$ plane of the solution at the end point of the $c=c(s)$ curve in Figure~\ref{fig:c(s);vor=0,-1}.
(right) For $\omega=-1$ and $d=1$, the wave profile of the solution at the end point of the solution curve in Figure~\ref{fig:c(s);vor=0,-1}. The mean fluid surface is at $y=0$ and the mean fluid depths are marked by dashed lines.}
\label{fig:y(x);vor=0,-1}
\end{figure}

We begin by taking $\omega=0$, and numerically solve \eqref{E:F=0} and \eqref{def:Y}-\eqref{def:B} using the method in the previous section. Figure~\ref{fig:c(s);vor=0,-1} includes the wave speed versus steepness from the result. It resembles the well-known result in an irrotational flow of infinite depth; see the inset of Figure~\ref{fig:c(s);d=infty}, for instance. 

The left panel of Figure~\ref{fig:y(x);vor=0,-1} displays the wave profile of the numerical solution at the end point of the $c=c(s)$ curve in Figure~\ref{fig:c(s);vor=0,-1}, in the $(x,y)$ plane for $x\in[-\upi,\upi]$, for which calculated are $c=0.9679$, $s=0.1024$ and $h=1.0398$. The mean fluid surface is at $y=0$. Clearly, it is near an {\em extreme wave}, which exhibits a sharp corner at the crest. Like in the infinite depth \citep[see][for instance]{LHF1978}, we conjecture that $c$ experiences infinitely many oscillations and $s$ increases monotonically toward the extreme wave. But the numerical solutions beyond the end point of the $c=c(s)$ curve in Figure~\ref{fig:c(s);vor=0,-1} require more than $2^{16}=65536$ Fourier coefficients to satisfy \eqref{E:error}, whence we do not compute them. We note that $h$ varies little throughout the $c=c(s)$ curve. 

For $\omega=-1$, in Figure~\ref{fig:c(s);vor=0,-1} is the wave speed versus steepness. It resembles the result for $\omega=0$. In the right panel of Figure~\ref{fig:y(x);vor=0,-1} is the wave profile of the numerical solution at the end point of the $c=c(s)$ curve; $c=0.6285$, $s=0.0404$ and $h=1.0286$. The steepness is noticeably less than for $\omega=0$. 

We verify what \cite{PTdS1988} obtained, using a boundary integral method. For instance, for $\omega=-1$ and $h=1$, \citep[Figure~$3(a)$]{PTdS1988} reports a solution for $c=0.5883$ and the crest-to-trough height $=0.12$. We find one for $c=0.5883$ and the crest-to-trough height $=0.1201$, $d=0.9978$. 

We may carry out the numerical computation for other values of negative vorticity. We predict that the result resembles that for $\omega=0$ or $-1$. See \cite{SS1985}, for instance, for some details, but for $d=\infty$. We predict that the crest becomes lower for stronger negative vorticity. 
 
\subsubsection*{Positive vorticity}

\begin{figure}
\centerline{\includegraphics[scale=1.1]{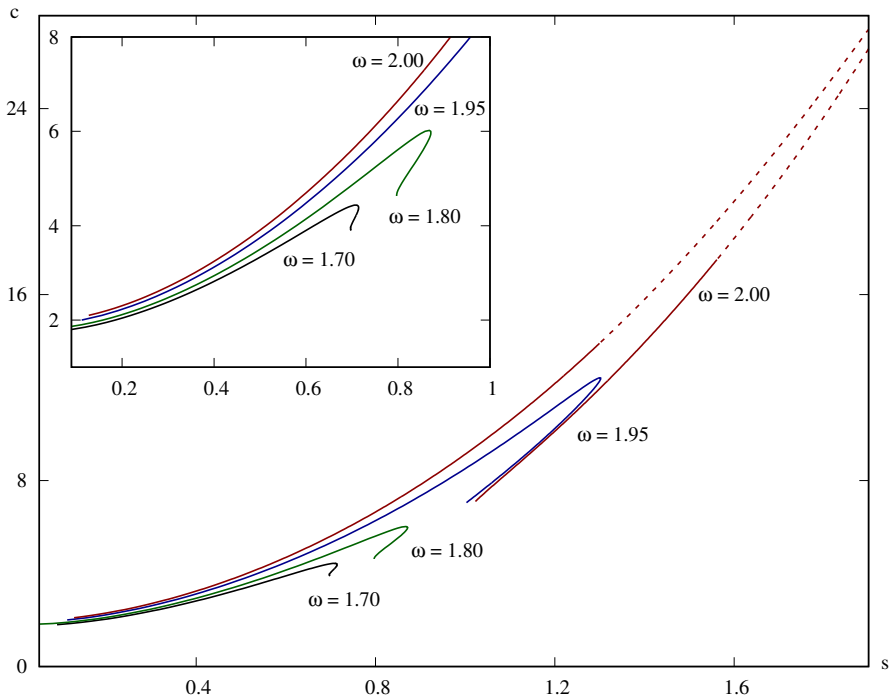}}
\caption{
Wave speed versus steepness for $d=1$, $\omega=1.7$ (black), $1.8$ (green), $1.95$ (blue) and $2$ (red). Solid curves physical solution, and the dashed curve unphysical solution. The inset is a closeup of the $c=c(s)$ curves for $s<1$.}
\label{fig:c(s);+vor;d=1}
\end{figure}

For weak positive vorticity, the result resembles that for zero or negative vorticity (see \cite{SS1985}, for instance, for some details for $d=\infty$), but not for strong positive vorticity. Figure~\ref{fig:c(s);+vor;d=1} includes the wave speed versus steepness for several values of positive vorticity.

For $\omega=1.7$, the inset of Figure~\ref{fig:c(s);+vor;d=1} reveals that the steepness increases, decreases and then increases along the $c=c(s)$ curve. Namely, a {\em fold} develops. Consequently, there correspond two or three solutions for some $s$. We predict that the continuation of the solution is limited by an extreme wave, which seems no longer the wave of greatest height. We note that there are no overhanging profiles throughout the $c=c(s)$ curve.

Figure~\ref{fig:c(s);+vor;d=1} indicates that the fold increases in size as $\omega$ increases. Moreover, waves observedly become more rounded for stronger positive vorticity. 
In particular, 
for $\omega=1.95$, we find an overhanging wave, whose profile is no longer the graph of a single valued function.

\begin{figure}
\centerline{\includegraphics[scale=1.1]{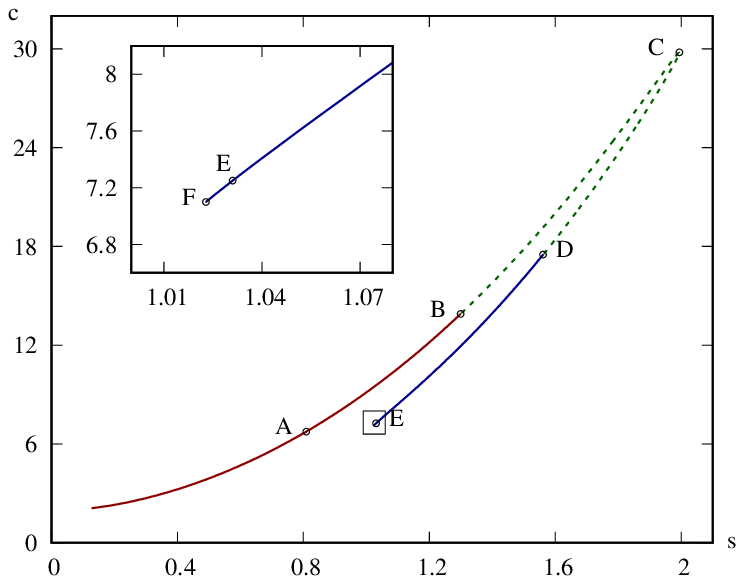}}
\centerline{\includegraphics[scale=1.1]{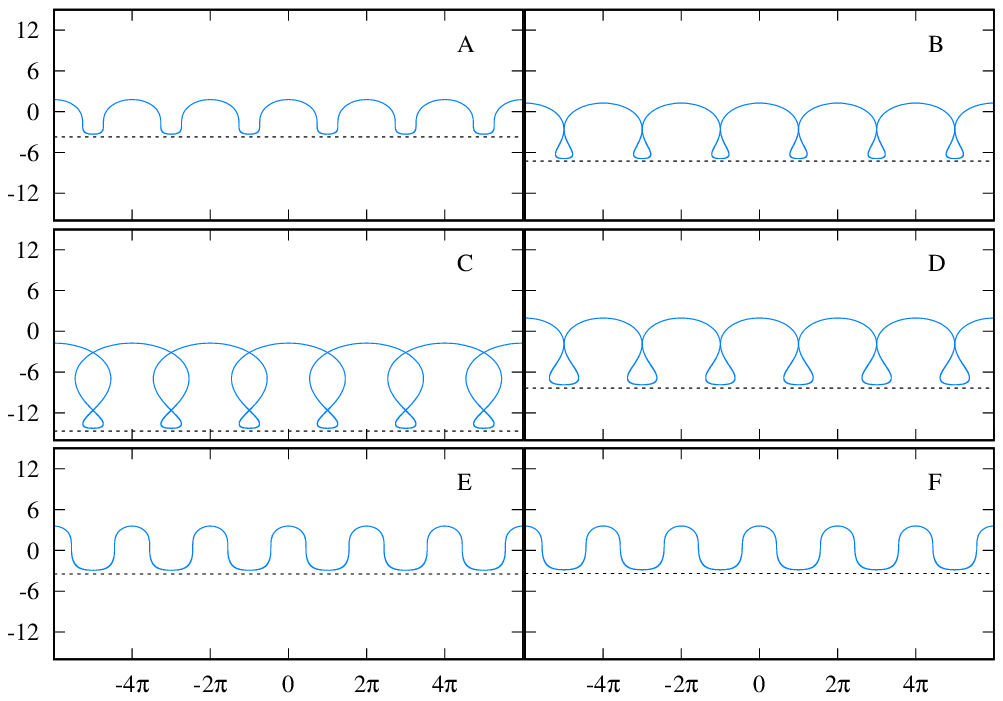}}
\caption{
(top) Wave speed versus steepness for $\omega=2$ and $d=1$. The inset is a closeup near the end point of the numerical continuation. (bottom) Wave profiles of the solutions labelled by $A$ through $F$. The mean fluid surface is at $y=0$ and the mean fluid depths are marked by dashed lines.}
\label{fig:c(s);vor=2}
\end{figure}

For $\omega=2$, the top panel of Figure~\ref{fig:c(s);vor=2} (see also Figure~\ref{fig:c(s);+vor;d=1}) shows the wave speed versus steepness. We continue the solution along the $c=c(s)$ curve from zero to a {\em touching wave}, whose profile self-intersects somewhere along the trough line, trapping an air bubble. Beyond such a limiting wave, the numerical solution becomes unphysical because the fluid surface in the conformal variable, $u\mapsto (u+\mathcal{T}_dy(u), y(u))$, (see \eqref{E:Ty}) is no longer injective for $u\in[-\upi,\upi]$. 
We continue the unphysical solution along the fold of the $c=c(s)$ curve, to reach another touching wave, and beyond such a limiting wave, the numerical solution becomes physical. Therefore, a {\em gap} of unphysical solutions develops in the $c=c(s)$ curve, which is limited by touching waves. We remark that the numerical method in \cite{SS1985}, for instance, diverges in the gap. In stark contrast, the present numerical method converges throughout the $c=c(s)$ curve. 

The bottom panel of Figure~\ref{fig:c(s);vor=2} displays six profiles along the $c=c(s)$ curve: $s=0.8092$, $h=3.6882$; $s=1.2997$, $h=7.2563$; $s=1.9939$, $h=1.4664$; $s=1.5615$, $h=8.3509$; $s=1.0310$, $h=3.4647$; $s=1.0228$, $h=3.3992$, respectively. 

Wave $A$ is single valued and $B$ is near a touching wave. By the way, it resembles a limiting capillary wave. Overhanging occurs somewhere between waves $A$ and $B$. Wave $C$ is in the gap, whose steepness is the maximum. It is unphysical because the fluid region over one period overlaps adjacent ones. Wave $D$ is near another touching wave. We find that wave $D$ traps a larger air bubble than $B$. Wave $E$ is single valued. We find that waves become more rounded from zero to wave $C$ and less rounded beyond $C$, so that overhanging ultimately disappears toward the extreme wave. Wave $F$ is at the end point of the $c=c(s)$ curve in the top panel, beyond which the numerical solutions require more than $2^{16}=65536$ Fourier coefficients for accurate resolution. We note that $h$ varies wildly along the $c=c(s)$ curve.

\begin{figure}
\centerline{\includegraphics[scale=1.1]{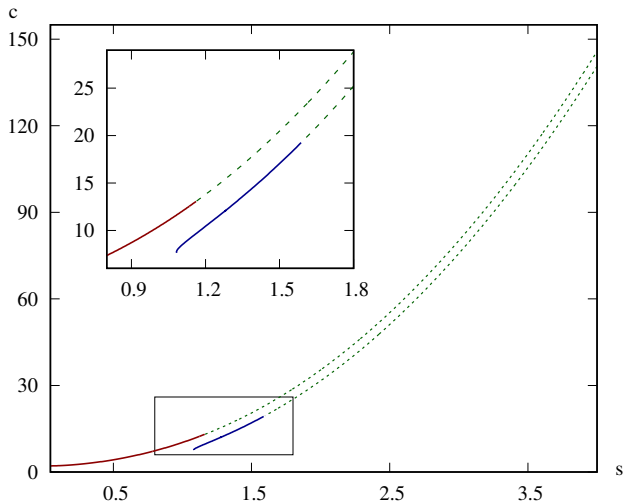}}
\caption{
Wave speed versus steepness for $\omega=2.25$ and $d=1$. The inset is a closeup near the limiting waves.}
\label{fig:c(s);vor=2.25}
\end{figure}

Moreover, we find that the gap increases in size as $\omega$ increases. For instance, for $\omega=2.25$, Figure~\ref{fig:c(s);vor=2.25} reveals that the wave speed reaches the order of hundreds in the gap. To compare, for $\omega=2$ (see Figure~\ref{fig:c(s);vor=2}), the wave speed does not exceed $30$. By the way, for $\omega=2.25$, we discontinue the numerical solution at $c=220$ and allow \eqref{E:res9} in place of \eqref{E:res13} in the gap. In order to locate a solution in the branch from a touching wave to an extreme wave of the $c=c(s)$ curve, without tracing out the gap, we begin by taking $\omega=2$ and a physical solution in the touching-to-extreme branch, and continue it along in $\omega$ toward $\omega=2.25$ while $c$ is held fixed. 

To summarize, the fold develops around $\omega=1.7$, and increases in size as $\omega$ increases. Overhanging waves appear for some $\omega$ in the range $1.8$ to $1.95$. The gap develops for some $\omega$ in the range $1.95$ to $2$, and becomes larger as $\omega$ increases. 

For stronger positive vorticity, we predict that one continues the solution from zero to a touching wave, which marks the outset of a gap; across the gap is another touching wave, which encloses a larger air bubble; one continues the solution through a fold until one reaches an extreme wave. 

\subsection{Infinite depth}\label{sec:infty-depth}

We turn the attention to $d=\infty$. For zero vorticity, \cite{LDS2017} proposed an auxiliary conformal mapping: for $\lambda>0$, 
\begin{equation}\label{def:zeta}
w(\zeta)=2\arctan\Big(\lambda\tan\frac{\zeta}{2}\Big),
\quad\text{where}\quad \zeta=\xi+i\eta\quad\text{and}\quad w=u+iv,
\end{equation}
conformally maps the lower half plane of $\mathbb{C}$ of $2\upi$ period in the $\xi$ variable to the lower half plane of $\mathbb{C}$ of $2\upi$ period in the $u$ variable and, moreover, $w\to\pm\upi$ as $\zeta\to\pm\upi$. 

Since $u\sim\lambda\xi$ for $\lambda\ll1$, \eqref{def:zeta} maps uniform grid points in the $\xi$ variable to non-uniform in the $u$ variable, concentrating them near $u=0$ by a factor of $\lambda$ and spreading them out near $u=\pm\upi$ by a factor of $1/\lambda$. Moreover, an FFT computes a discrete Fourier transform in $O(\sqrt{N})$ operations in the $\xi$ variable, whereas it takes $O(N)$ operations in the $u$ variable, where $N$ is the number of grid points over one period. It is particularly effective for nearly extreme waves, whose crests tend to sharpen and troughs flatter. For instance, in an irrotational flow of infinite depth, a uniform grid in the $u$ variable requires $2^{27}\approx1.3\times10^8$ Fourier coefficients to estimate the wave of greatest height up to $32$ digits \citep[see][for instance]{DLK2016}, whereas a uniform grid in the $\xi$ variable and, hence, a non-uniform grid in the $u$ variable use about $4.2\times10^4$ Fourier coefficients \citep[see][]{LDS2017} for comparable resolution.

The result in the subsection makes use of \eqref{def:zeta} for nonzero constant vorticity. Unfortunately, to the best of the authors' knowledge, no such mapping is known in the finite depth. It is an interesting question for future investigation.

\begin{figure}
\centerline{\includegraphics[scale=1.1]{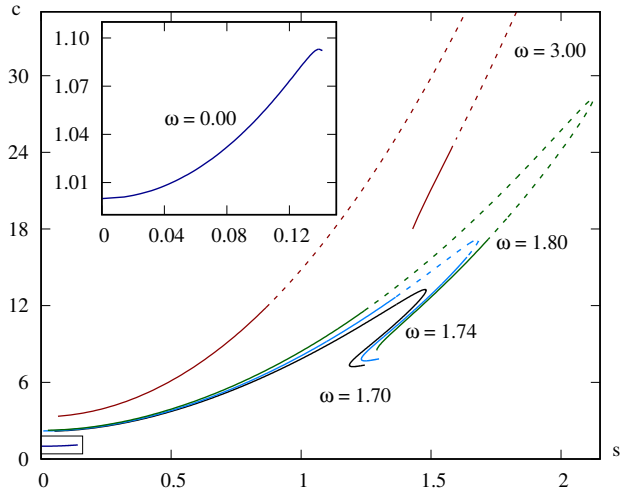}}
\caption{
Wave speed versus steepness for $d=\infty$, $\omega = 0$ (blue), $1.7$ (black), $1.8$ (green) and $3$ (red). Solid curves for physical solution, and dashed curves for unphysical solution. The inset is a closeup of the $c=c(s)$ curve for $\omega=0$. 
}
\label{fig:c(s);d=infty}
\end{figure}

\begin{figure}
\centerline{\includegraphics[scale=1.1]{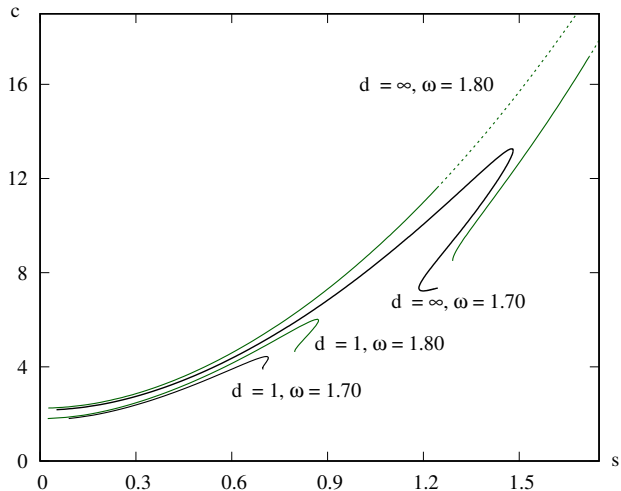}}
\caption{Wave speed versus steepness for $d=1$ (green) and $\infty$ (black) for $\omega = 1.7$ and $1.8$.
}
\label{fig:d-comparison}
\end{figure}

Figure~\ref{fig:c(s);d=infty} includes the wave speed versus steepness for several values of vorticity. It agrees with \citep[Figure~$9$]{SS1985}, using a boundary integral method. By the way, the vorticity in \cite{SS1985} differs in sign. Moreover, it resembles the result in the finite depth; see Figure~\ref{fig:c(s);+vor;d=1}. Indeed, Figure~\ref{fig:d-comparison} indicates that the effects of depth are merely to change steepness and other quantities, and they are insignificant otherwise. We note that greater depths allow larger waves for higher speeds.

For $\omega=0$, the inset of Figure~\ref{fig:c(s);d=infty} reproduces the well-known result of \cite{LHF1978}, among others, that single valued profiles tend to an extreme wave as the steepness increases monotonically. See also \cite{LDS2017}, among others. Like in the previous subsection in the finite depth, the fold develops for some $\omega$ in the range $1.3$ to $1.5$, and increases in size as $\omega$ increases. Overhanging waves appear for some $\omega$ in the range $1.6$ to $1.7$. The gap develops around $\omega=1.7$, and becomes larger as $\omega$ increases. See \cite{SS1985} for more details. 

For $\omega=3$, there seems to correspond two, one or zero solutions for some $s$ less than the maximum. By the way, like in the previous subsection in the finite depth, we discontinue the numerical solution at $c$ at the order of hundreds, and locate a solution in the touching-to-extreme branch of the $c=c(s)$ curve by continuing along in $\omega$ from a smaller value. 

\begin{figure}
\centerline{\includegraphics[scale=1.1]{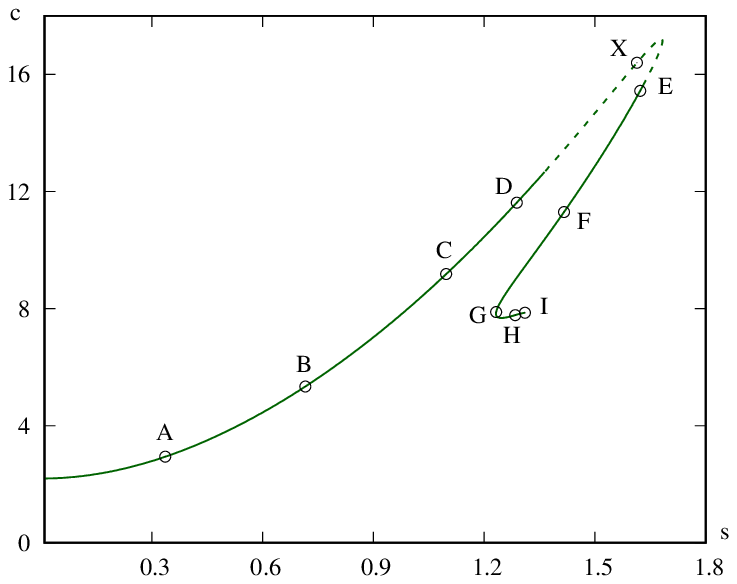}}
\centerline{\includegraphics[scale=1.1]{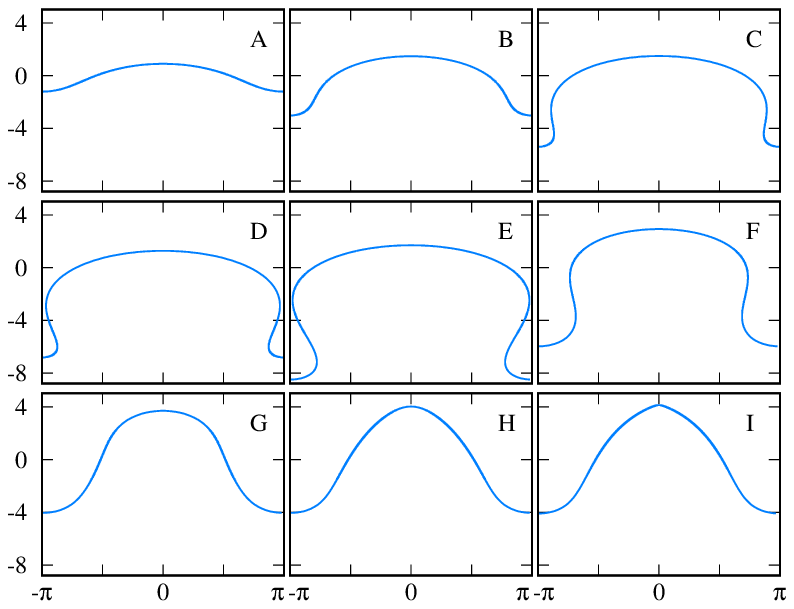}}
\centerline{\includegraphics[scale=1.0]{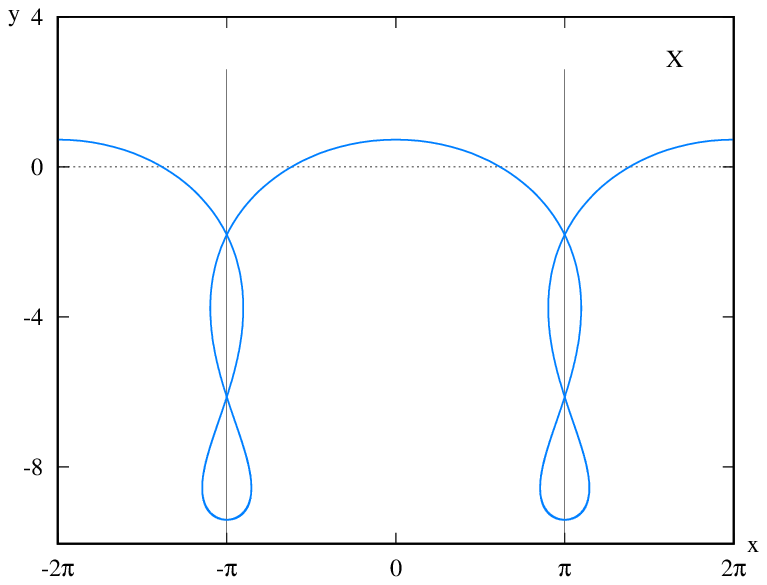}}
\caption{(top) Wave speed versus steepness for $\omega=1.74$ and $d=\infty$. (middle) Wave profiles of the solutions labelled by $A$ through $I$. The mean fluid surface is at $y=0$. (bottom) Wave profile of the solution labelled by $X$ in Figure~\ref{fig:simmen}.}
\label{fig:simmen}
\end{figure}

Figure~\ref{fig:simmen} displays  the wave speed versus steepness for $\omega=1.74$ and a selection of wave profiles along the $c=c(s)$ curve. The middle panel is, qualitatively and quantitatively, in excellent agreement with \citep[Figure~$8$]{SS1985}. The bottom panel shows the wave profile of the numerical solution in the gap of the $c=c(s)$ curve, labelled by $X$, for which calculated are $c=16.4$ and $s=1.6139$. Clearly, it is unphysical. By the way, the numerical method in \cite{SS1985}, for instance, diverges in the gap.

\subsection{Large vorticity limit}\label{sec:max omega}

Throughout the subsection, $d=\infty$.

\begin{figure}
\centerline{\includegraphics[scale=1.1]{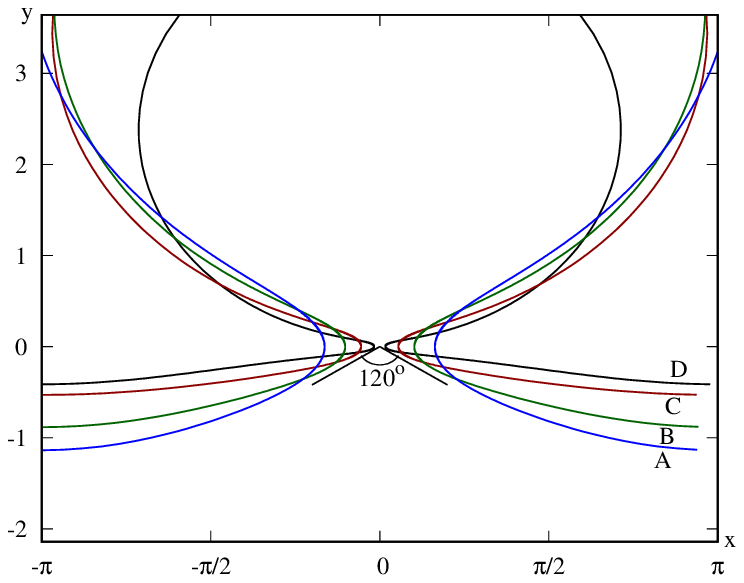}}
\caption{Wave profiles of the solutions for $d=\infty$, $\omega=4.3$ and $c =34.0$ ($A$, blue), $\omega=5.5$ and $c= 36.0$ ($B$, green), $\omega=7.6$ and $c = 48.5$ ($C$, red), and $\omega=14.0$ and $c=51.7$ ($D$, black).}
\label{fig:vor->infty}
\end{figure}

In Figure~\ref{fig:vor->infty}, we begin by taking $\omega=4.3$ and wave $A$ in the gap close to the touching-to-extreme branch of the $c=c(s)$ curve, for which $c=34.0$. We take wave $A$ as the initial guess and continue the solution along in $\omega$ and $c$, to reach wave $B$, for which $\omega=5.5$ and $c=36.0$. We continue the solution along in $\omega$ and $c$, likewise, to reach waves $C$ and $D$; $\omega=7.6$, $c=48.5$ and $\omega=14.0$, $c=51.7$, respectively. We note that waves $B$, $C$ and $D$ are in the touching-to-extreme branches of the $c=c(s)$ curves. They become more rounded for stronger vorticity, and the trough becomes wider. Consequently, a {\em neck} develops in the profile, which decreases in size as $\omega$ increases. We note that waves $B$, $C$ and $D$ resemble the profiles in \citep[Figure~$5$][for instance]{VB1996} at the zero gravity limit.

\begin{figure}
\centerline{\includegraphics[scale=1.1]{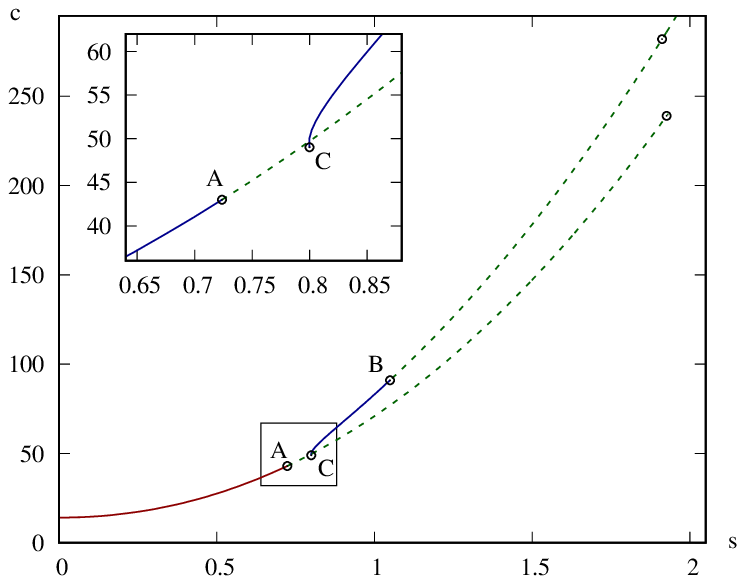}}
\centerline{\includegraphics[scale=1.1]{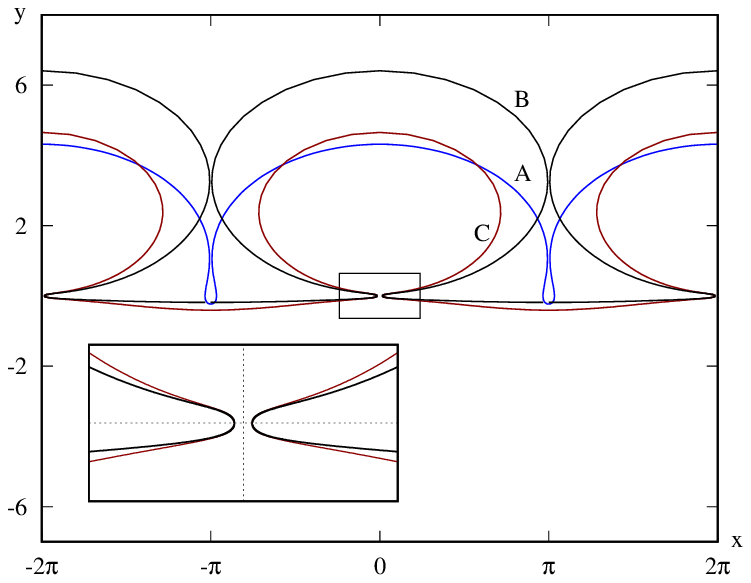}}
\caption{(top) Wave speed versus steepness for $\omega=14$ and $d=\infty$. Solid curves for physical solution and dashed curves for unphysical solution. The inset is a closeup near the boundaries of the two solution branches. (bottom) Wave profiles of the solutions labelled by $A$, $B$ and $C$. The inset is a closeup near the neck.} 
\label{fig:w14}
\end{figure}

Figure~\ref{fig:w14} displays the wave speed versus steepness for $\omega=14$ and a selection of wave profiles along the $c=c(s)$ curve. Wave $A$ is near the end point of the zero-to-touching branch of the $c=c(s)$ curve. Wave $B$ is near the outset of the touching-to-extreme branch, and wave $C$ is the end point of the $c=c(s)$ curve. Beyond wave $C$, the numerical solutions require more than $2^{16}=65536$ Fourier coefficients for accurate resolution. Interestingly, the touching-to-extreme branch seems to intersect the gap. Wave $D$ of Figure~\ref{fig:vor->infty} seems located somewhere between waves $B$ and $C$ of Figure~\ref{fig:w14}. By the way, we discontinue the numerical solution in the gap at $c\approx250$, and locate wave $B$ by continuing along in $\omega$ and $c$ from smaller values. 

We note that wave $A$ closely resembles \citep[Figure~$4(b)$][for instance]{VB1996} at the zero gravity limit. \cite{DH2} will study in detail the limiting wave at the end points of the zero-to-touching branches of the $c=c(s)$ curves as the strength of positive vorticity increases unboundedly or, equivalently, as gravitational acceleration vanishes. 

We find that the neck decreases in size along the gap until it reaches a minimum, for which $c$ is much less than a maximum, and then remains constant, particularly, from waves $B$ to $C$. On the other hand, we note that waves become more rounded along the fold until $c$ reaches the maximum, and then less rounded. Moreover, we note that the numerical solutions are ultimately limited by an extreme wave, which exhibits a sharp corner at the crest. Together, we predict that the ``fluid bubble" disappears somewhere in the touching-to-extreme branch of the $c=c(s)$ curve. 

Figure~\ref{fig:vor->infty} reveals that the neck becomes narrower for stronger positive vorticity. 
Indeed, we predict that the minimum neck size vanishes as the strength of positive vorticity increases unboundedly. Moreover, we predict that the limiting wave at the large vorticity limit is rigid body rotation of a fluid disk. \cite{DH2} will confirm it.

\subsection{Minimum depth limit}\label{sec:min depth}

\begin{figure}
\centerline{\includegraphics[scale=1]{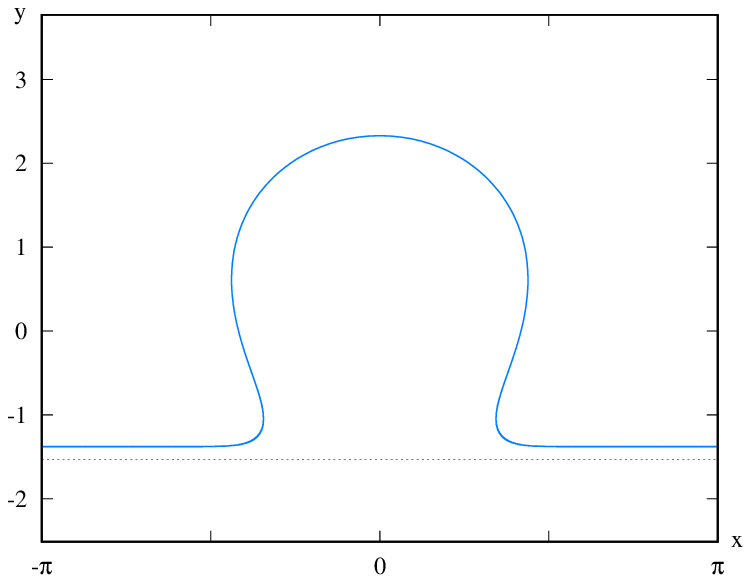}}
\caption{For $\omega=3$ and $c=6$, the wave profile near the minimum of the mean conformal depth. The mean fluid surface is $y=0$ and the mean fluid depth is marked by the dashed line.
} 
\label{fig4:min-depth}
\end{figure}

Lastly, we investigate a limit as the mean conformal depth decreases to the minimum. For instance, for $\omega=3$ and $c=6$ fixed, $d$ decreases to $\approx0.2193$. Figure~\ref{fig4:min-depth} displays the wave profile of the numerical solution near such a minimum value, for which calculated is $h=1.5329$. It resembles \citep[Figure~7(a)]{PTdS1988}, for instance, whose trough is flat and limited by the fluid bed. 

\

\subsection*{Acknowledgements}
VMH is supported by the National Science Foundation under the Faculty Early Career Development (CAREER) Award DMS-1352597, an Alfred P. Sloan Research Fellowship, a Simons Fellowship in Mathematics, and by the University of Illinois at Urbana-Champaign under the Arnold O. Beckman Research Award RB14100. She is grateful to the Department of Mathematics at Brown University for its generous hospitality.
SD is supported by the National Science Foundation Award DMS-1716822.

This material is based on work supported by the National Science Foundation under DMS-1439786 while the authors were in residence at the Institute for Computational and Experimental Research in Mathematics in Providence, RI, during the Spring 2017 semester.

\

\bibliographystyle{jfm}
\bibliography{vorticitybib}

\begin{thebibliography}{42}
\expandafter\ifx\csname natexlab\endcsname\relax\def\natexlab#1{#1}\fi
\def\au#1{#1} \def\ed#1{#1} \def\yr#1{#1}\def\at#1{#1}\def\jt#1{\textit{#1}}
  \def\bt#1{#1}\def\bvol#1{\textbf{#1}} \def\vol#1{#1} \def\pg#1{#1}
  \def\publ#1{#1}\def\arxiv#1{#1}\def\org#1{#1}\def\st#1{\textit{#1}}

\bibitem[Amick {\em et~al.\/}(1982)Amick, Fraenkel \& Toland]{AFT1982}
{\sc \au{Amick, C.~J.}, \au{Fraenkel, L.~E.} \& \au{Toland, J.~F.}} \yr{1982}
  \at{On the {S}tokes conjecture for the wave of extreme form}.  \jt{Acta
  Math.}  \bvol{148},  \pg{193--214}.

\bibitem[Babenko(1987)]{Babenko1987}
{\sc \au{Babenko, K.~I.}} \yr{1987}  \at{Some remarks on the theory of surface
  waves of finite amplitude}.  \jt{Soviet Math. Doklady}  \bvol{35}~(6),
  \pg{599--603}, ({S}ee also {\em loc. cit.} 647--650).

\bibitem[Buffoni {\em et~al.\/}(2000{\natexlab{{\em a\/}}})Buffoni, Dancer \&
  Toland]{BDT2000a}
{\sc \au{Buffoni, B.}, \au{Dancer, E.~N.} \& \au{Toland, J.~F.}}
  \yr{2000{\natexlab{{\em a\/}}}}  \at{The regularity and local bifurcation of
  steady periodic water waves}.  \jt{Arch. Ration. Mech. Anal.}
  \bvol{152}~(3),  \pg{207--240}.

\bibitem[Buffoni {\em et~al.\/}(2000{\natexlab{{\em b\/}}})Buffoni, Dancer \&
  Toland]{BDT2000b}
{\sc \au{Buffoni, B.}, \au{Dancer, E.~N.} \& \au{Toland, J.~F.}}
  \yr{2000{\natexlab{{\em b\/}}}}  \at{The sub-harmonic bifurcation of {S}tokes
  waves}.  \jt{Arch. Ration. Mech. Anal.}  \bvol{152}~(3),  \pg{241--271}.

\bibitem[Constantin {\em et~al.\/}(2007)Constantin, Ehrnstr\"om \&
  Wahl\'en]{CEW2007}
{\sc \au{Constantin, Adrian}, \au{Ehrnstr\"om, Mats} \& \au{Wahl\'en, Erik}}
  \yr{2007}  \at{Symmetry of steady periodic gravity water waves with
  vorticity}.  \jt{Duke Math. J.}  \bvol{140}~(3),  \pg{591--603}.

\bibitem[Constantin \& Strauss(2004)]{CS2004}
{\sc \au{Constantin, Adrian} \& \au{Strauss, Walter}} \yr{2004}  \at{Exact
  steady periodic water waves with vorticity}.  \jt{Comm. Pure Appl. Math.}
  \bvol{57}~(4),  \pg{481--527}.

\bibitem[Constantin {\em et~al.\/}(2016)Constantin, Strauss \&
  V\u{a}rv\u{a}ruc\u{a}]{CSV2016}
{\sc \au{Constantin, Adrian}, \au{Strauss, Walter} \&
  \au{V\u{a}rv\u{a}ruc\u{a}, Eugen}} \yr{2016}  \at{Global bifurcation of
  steady gravity water waves with critical layers}.  \jt{Acta Math.}
  \bvol{217}~(2),  \pg{195--262}.

\bibitem[Dyachenko {\em et~al.\/}(1996{\natexlab{{\em a\/}}})Dyachenko,
  Kuznetsov, Spector \& Zakharov]{DKSZ1996}
{\sc \au{Dyachenko, A.~I.}, \au{Kuznetsov, E.~A.}, \au{Spector, M.~D.} \&
  \au{Zakharov, V.~E.}} \yr{1996{\natexlab{{\em a\/}}}}  \at{Analytical
  description of the free surface dynamics of an ideal fluid (canonical
  formalism and conformal mapping)}.  \jt{Physics Letters A}  \bvol{221}~(1),
  \pg{73--79}.

\bibitem[Dyachenko {\em et~al.\/}(1996{\natexlab{{\em b\/}}})Dyachenko,
  Zakharov \& Kuznetsov]{DZK1996}
{\sc \au{Dyachenko, A.~I.}, \au{Zakharov, V.~E.} \& \au{Kuznetsov, E.~A.}}
  \yr{1996{\natexlab{{\em b\/}}}}  \at{Nonlinear dynamics of the free surface
  of an ideal fluid}.  \jt{Plasma Physics Reports}  \bvol{22}~(10),
  \pg{829--840}.

\bibitem[Dyachenko \& Hur(2018)]{DH2}
{\sc \au{Dyachenko, Sergey~A.} \& \au{Hur, Vera~Mikyoung}} \yr{2018}
  \at{Stokes waves with constant vorticity: {II}. limiting waves}.
  \jt{Preprint} .

\bibitem[Dyachenko {\em et~al.\/}(2015)Dyachenko, Lushnikov \&
  Korotkevich]{DLK}
{\sc \au{Dyachenko, S.~A.}, \au{Lushnikov, P.~M.} \& \au{Korotkevich, A.~O.}}
  \yr{2015} Library of {S}tokes waves. Published electronically at
  {http://stokeswave.org}.

\bibitem[Dyachenko {\em et~al.\/}(2016)Dyachenko, Lushnikov \&
  Korotkevich]{DLK2016}
{\sc \au{Dyachenko, S.~A.}, \au{Lushnikov, P.~M.} \& \au{Korotkevich, A.~O.}}
  \yr{2016}  \at{Branch cuts of {S}tokes wave on deep water. {P}art {I}:
  {N}umerical solution and {P}ad\'e approximation}.  \jt{Stud. Appl. Math.}
  \bvol{137}~(4),  \pg{419--472}.

\bibitem[Gakhov(1990)]{Gakhov}
{\sc \au{Gakhov, F.~D.}} \yr{1990} {\em Boundary Value Problems\/}.
  \publ{Dover Publications, Inc., New York}, translated from the Russian,
  Reprint of the 1966 translation.

\bibitem[Greenbaum(1997)]{Greenbaum1997}
{\sc \au{Greenbaum, Anne}} \yr{1997} {\em Iterative Methods for Solving Linear
  Systems\/},  \st{Frontiers in Applied Mathematics},  \vol{vol.~17}.
  \publ{Society for Industrial and Applied Mathematics (SIAM), Philadelphia,
  PA}.

\bibitem[Hur(2006)]{Hur2006}
{\sc \au{Hur, Vera~Mikyoung}} \yr{2006}  \at{Global bifurcation theory of
  deep-water waves with vorticity}.  \jt{SIAM J. Math. Anal.}  \bvol{37}~(5),
  \pg{1482--1521}.

\bibitem[Hur(2007)]{Hur2007}
{\sc \au{Hur, Vera~Mikyoung}} \yr{2007}  \at{Symmetry of steady periodic water
  waves with vorticity}.  \jt{Philos. Trans. R. Soc. Lond. Ser. A Math. Phys.
  Eng. Sci.}  \bvol{365}~(1858),  \pg{2203--2214}.

\bibitem[Hur(2011)]{Hur2011}
{\sc \au{Hur, Vera~Mikyoung}} \yr{2011}  \at{Stokes waves with vorticity}.
  \jt{J. Anal. Math.}  \bvol{113},  \pg{331--386}.

\bibitem[Ko \& Strauss(2008{\natexlab{{\em a\/}}})]{KS1}
{\sc \au{Ko, Joy} \& \au{Strauss, Walter}} \yr{2008{\natexlab{{\em a\/}}}}
  \at{Effect of vorticity on steady water waves}.  \jt{J. Fluid Mech.}
  \bvol{608},  \pg{197--215}.

\bibitem[Ko \& Strauss(2008{\natexlab{{\em b\/}}})]{KS2}
{\sc \au{Ko, Joy} \& \au{Strauss, Walter}} \yr{2008{\natexlab{{\em b\/}}}}
  \at{Large-amplitude steady rotational water waves}.  \jt{Eur. J. Mech. B
  Fluids}  \bvol{27}~(2),  \pg{96--109}.

\bibitem[Longuet-Higgins(1978)]{LH1978}
{\sc \au{Longuet-Higgins, M.~S.}} \yr{1978}  \at{Some new relations between
  {S}tokes's coefficients in the theory of gravity waves}.  \jt{J. Inst. Math.
  Appl.}  \bvol{22}~(3),  \pg{261--273}.

\bibitem[Longuet-Higgins \& Fox(1978)]{LHF1978}
{\sc \au{Longuet-Higgins, M.~S.} \& \au{Fox, M. J.~H.}} \yr{1978}  \at{Theory
  of the almost-highest wave. {II}. {M}atching and analytic extension}.  \jt{J.
  Fluid Mech.}  \bvol{85}~(4),  \pg{769--786}.

\bibitem[Lushnikov(2016)]{Lushnikov2016}
{\sc \au{Lushnikov, Pavel~M.}} \yr{2016}  \at{Structure and location of branch
  point singularities for {S}tokes waves on deep water}.  \jt{J. Fluid Mech.}
  \bvol{800},  \pg{557--594}.

\bibitem[Lushnikov {\em et~al.\/}(2017)Lushnikov, Dyachenko \&
  Silantyev]{LDS2017}
{\sc \au{Lushnikov, Pavel~M.}, \au{Dyachenko, Sergey~A.} \& \au{Silantyev,
  Denis~A.}} \yr{2017}  \at{New conformal mapping for adaptive resolving of the
  complex singularities of {S}tokes wave}.  \jt{Proc. A.}  \bvol{473}~(2202),
  \pg{20170198, 19}.

\bibitem[Meiron {\em et~al.\/}(1981)Meiron, Orszag \& Israeli]{MOI1981}
{\sc \au{Meiron, Daniel~I.}, \au{Orszag, Steven~A.} \& \au{Israeli, Moshe}}
  \yr{1981}  \at{Applications of numerical conformal mapping}.  \jt{J. Comput.
  Phys.}  \bvol{40}~(2),  \pg{345--360}.

\bibitem[Meurant(1999)]{Meurant1999}
{\sc \au{Meurant, G.}} \yr{1999} {\em Computer Solution of Large Linear
  Systems\/},  \st{Studies in Mathematics and its Applications},
  \vol{vol.~28}.  \publ{North-Holland Publishing Co., Amsterdam}.

\bibitem[Ovsyannikov(1973)]{Ovsyannikov1973}
{\sc \au{Ovsyannikov, Lev~V}} \yr{1973}  \at{Dynamics of a fluid}.  \jt{MA
  Lavrent{'}ev Institute of Hydrodynamics Sib. Branch USSR Ac. Sci}  \bvol{15},
   \pg{104--125}.

\bibitem[Plemelj(1964)]{Plemelj}
{\sc \au{Plemelj, Josip}} \yr{1964} {\em Problems in the Sense of {R}iemann and
  {K}lein\/}.  \publ{Interscience Publishers John Wiley \& Sons Inc.\, New
  York-London-Sydney}.

\bibitem[Plotnikov(1992)]{Plotnikov1992}
{\sc \au{Plotnikov, P.~I.}} \yr{1992}  \at{Nonuniqueness of solutions of the
  problem of solitary waves and bifurcation of critical points of smooth
  functionals}.  \jt{Math. USSR Izvestiya}  \bvol{38}~(2),  \pg{333--357}.

\bibitem[Ribeiro {\em et~al.\/}(2017)Ribeiro, Milewski \& Nachbin]{RMN2017}
{\sc \au{Ribeiro, Jr., Roberto}, \au{Milewski, Paul~A.} \& \au{Nachbin,
  Andr\'e}} \yr{2017}  \at{Flow structure beneath rotational water waves with
  stagnation points}.  \jt{J. Fluid Mech.}  \bvol{812},  \pg{792--814}.

\bibitem[Saad(2003)]{Saad2003}
{\sc \au{Saad, Yousef}} \yr{2003} {\em Iterative Methods for Sparse Linear
  Systems\/}, 2nd edn.  \publ{Society for Industrial and Applied Mathematics,
  Philadelphia, PA}.

\bibitem[Saad \& Schultz(1986)]{SS1986}
{\sc \au{Saad, Youcef} \& \au{Schultz, Martin~H.}} \yr{1986}  \at{G{MRES}: a
  generalized minimal residual algorithm for solving nonsymmetric linear
  systems}.  \jt{SIAM J. Sci. Statist. Comput.}  \bvol{7}~(3),  \pg{856--869}.

\bibitem[Teles~da Silva \& Peregrine(1988)]{PTdS1988}
{\sc \au{Teles~da Silva, A.~F.} \& \au{Peregrine, D.~H.}} \yr{1988}  \at{Steep,
  steady surface waves on water of finite depth with constant vorticity}.
  \jt{J. Fluid Mech.}  \bvol{195},  \pg{281--302}.

\bibitem[Simmen \& Saffman(1985)]{SS1985}
{\sc \au{Simmen, J.~A.} \& \au{Saffman, P.~G.}} \yr{1985}  \at{Steady
  deep-water waves on a linear shear current}.  \jt{Stud. Appl. Math.}
  \bvol{73}~(1),  \pg{35--57}.

\bibitem[Simoncini \& Szyld(2007)]{SS2007}
{\sc \au{Simoncini, Valeria} \& \au{Szyld, Daniel~B.}} \yr{2007}  \at{Recent
  computational developments in {K}rylov subspace methods for linear systems}.
  \jt{Numer. Linear Algebra Appl.}  \bvol{14}~(1),  \pg{1--59}.

\bibitem[Stokes(1847)]{Stokes1847}
{\sc \au{Stokes, G.~G.}} \yr{1847}  \at{On the theory of oscillatory waves}.
  \jt{Trans. Camb. Philos. Soc.}  \bvol{8},  \pg{441--473}.

\bibitem[Stokes(1880)]{Stokes1880}
{\sc \au{Stokes, G.~G.}} \yr{1880} {\em Mathematical and Physical Papers\/}, ,
  \vol{vol.~1}.  \publ{Cambridge University Press}.

\bibitem[Tanveer(1991)]{Tanveer1991}
{\sc \au{Tanveer, S.}} \yr{1991}  \at{Singularities in water waves and
  {R}ayleigh-{T}aylor instability}.  \jt{Proc. Roy. Soc. London Ser. A}
  \bvol{435}~(1893),  \pg{137--158}.

\bibitem[Tanveer(1993)]{Tanveer1993}
{\sc \au{Tanveer, S.}} \yr{1993}  \at{Singularities in the classical
  {R}ayleigh-{T}aylor flow: formation and subsequent motion}.  \jt{Proc. Roy.
  Soc. London Ser. A}  \bvol{441}~(1913),  \pg{501--525}.

\bibitem[Titchmarsh(1986)]{Titchmarsh}
{\sc \au{Titchmarsh, E.~C.}} \yr{1986} {\em Introduction to the Theory of
  {F}ourier Integrals\/}, 3rd edn.  \publ{Chelsea Publishing Co., New York}.

\bibitem[Vanden-Broeck(1996)]{VB1996}
{\sc \au{Vanden-Broeck, J.-M.}} \yr{1996}  \at{Periodic waves with constant
  vorticity in water of infinite depth}.  \jt{IMA J. Appl. Math.}
  \bvol{56}~(2),  \pg{207--217}.

\bibitem[Yang(2010)]{Yang2010}
{\sc \au{Yang, Jianke}} \yr{2010} {\em Nonlinear Waves in Integrable and
  Nonintegrable Systems\/},  \st{Mathematical Modeling and Computation},
  \vol{vol.~16}.  \publ{Society for Industrial and Applied Mathematics (SIAM),
  Philadelphia, PA}.

\bibitem[Zakharov {\em et~al.\/}(2002)Zakharov, Dyachenko \&
  Vasilyev]{ZDAV2002}
{\sc \au{Zakharov, Vladimir~E.}, \au{Dyachenko, Alexander~I.} \& \au{Vasilyev,
  Oleg~A.}} \yr{2002}  \at{New method for numerical simulation of a
  nonstationary potential flow of incompressible fluid with a free surface}.
  \jt{Eur. J. Mech. B Fluids}  \bvol{21}~(3),  \pg{283--291}.

\end{thebibliography}

\end{document}